\documentclass[12pt,twoside]{amsart}

\usepackage{amssymb}
\usepackage[mathscr]{eucal}

\newtheorem{prop}{Proposition}
\theoremstyle{definition}

\newcommand{\ds}{\displaystyle}

\def\EXP{\mbox{{\large\bf e}}}
\newcommand{\uop}{\mathbf{u}}
\newcommand{\wop}{\mathbf{w}}
\newcommand{\xop}{\mathbf{x}}
\newcommand{\zop}{\mathbf{z}}

\newcommand{\R}{\mathbf{R}}
\newcommand{\Q}{\mathbf{Q}}

\newcommand{\ltoda}{\ell}
\newcommand{\ldst}{{\widetilde{\ell}}}
\newcommand{\montoda}{\widehat{\mathbf{t}}}
\newcommand{\tp}{\mathbf{t}}
\newcommand{\Ltoda}{L}
\newcommand{\Ldst}{\widetilde{L}}
\newcommand{\Montoda}{\widehat{T}}

\newcommand{\Om}{\Delta}
\newcommand{\omeg}{{\omega^{1/2}}}
\newcommand{\omegg}{{\omega^{-1/2}}}
\newcommand{\ff}{\phi}

\newcommand{\Set}{\mathfrak{F}_M}
\def\Box{\blacksquare}

\begin{document}

\hfill ITEP-TH-24/02
\vspace{2cm}

\title[]{Quantum relativistic Toda chain at root of unity:
isospectrality, modified $Q$-operator and\\
functional Bethe ansatz}%
\author{S. Pakuliak}
\address{Stanislav Pakuliak, Bogoliubov Laboratory of Theoretical Physics, Joint Institute for Nuclear Research,
Dubna 141980, Moscow reg., Russia}%
\email{pakuliak@thsun1.jinr.ru}%
\author{S. Sergeev}%
\address{Sergei Sergeev, Bogoliubov Laboratory of Theoretical Physics, Joint Institute for Nuclear Research,
Dubna 141980, Moscow reg., Russia} \email{sergeev@thsun1.jinr.ru}

\thanks{This work was supported in part by the grant INTAS OPEN 00-00055.
S.P.'s work was supported by the grant RFBR 00-02-16477 and grant
for support of scientific schools RFBR 00-15-96557, S.S's work
was supported by the grant RFBR 01-01-00201.}%


%\date{July 18, 2001}% %\dedicatory{}% %\commby{}%
%-----------------------------------------------------
\begin{abstract}
We investigate an $N$-state spin model called quantum relativistic
Toda chain and based on  the unitary finite dimensional representations of
the Weyl algebra with $q$ being    $N$th primitive root of unity. Parameters of
the finite dimensional representation of the local Weyl algebra
form the classical discrete integrable system.  Nontrivial
dynamics of the classical counterpart corresponds to
isospectral transformations of the spin system. Similarity
operators are constructed with the help of modified Baxter's
$Q$-operators. The classical
counterpart of the modified $Q$-operator for the initial
homogeneous spin chain is a B\"acklund
transformation. This transformation creates an extra Hirota-type soliton in a
parameterization of the chain structure. Special choice of values
of solitonic amplitudes yields a degeneration of spin eigenstates,
leading to the quantum separation of variables, or the functional Bethe
ansatz. A projector to the separated eigenstates is constructed
explicitly as a product of modified $Q$-operators.
\end{abstract}

\maketitle

%------------------------------------------------------

%\newpage

%\tableofcontents

%\newpage

\section*{Introduction}

One of the indicative examples of integrable models of
mathematical physics is the Toda chain. There are an enormous
number of papers in the modern literature concerning the Toda
chain, classical as well as quantum one. Let us mention only ref.
\cite{Sklyanin-sep} as an example of the application of the algebraic methods
to the quantum Toda chain. One of the
modifications of the Toda chain is known as a relativistic
Toda chain, classical as well as quantum, see. e.g.
\cite{Ruijsenaars,Suris,Kundu,Kuznetsov,PS}.

The algebra of observables in the quantum relativistic Toda
chain is the local Weyl algebra. Due to an ambiguity of
centers of the Weyl algebra with arbitrary Weyl's $q$, one
may talk about well defined \emph{quantum} model only in three
cases. The formal first one is the special limit $q\mapsto 1$,
corresponding to the usual quantum Toda chain \cite{Sklyanin-sep}.
The  second case is
the modular dualization of Weyl's algebra, see
\cite{Faddeev-94,Faddeev-99,fkv} for details and \cite{kls} for
the application of the modular concept to the quantum relativistic
Toda chain. In this paper we are going to discuss the third known
case -- the case of the \emph{unitary finite dimensional}
representation of the Weyl algebra, arising when Weyl's parameter $q$
is the $N$th primitive root of unity. Note, besides the finite
dimensional representations at the root of unity, there exist the
\emph{real infinite dimensional} ones, but this case belongs to the
modular double class.

In several integrable models with the unitary representations of
the Weyl algebra at the $N$th root of unity, the
$\mathbb{C}$-valued $N$th powers of Weyl's elements form a
classical discrete integrable system. So the parameters of the
unitary representation of the Weyl algebra form a classical
counterpart of the finite dimensional integrable system (i.e. spin
integrable system), see
\cite{FK-qd,BR-qd,BR-unpublished,bbr,Sergeev} for examples. The
most important finite dimensional operators, arising in the spin
systems, have functional counterparts, defined as rational
mappings in the space of functions of the $N$th powers of Weyl's
elements. So the finite dimensional operators are the secondary
objects: one has to define first the mapping of the parameters,
and then the finite dimensional operators have to be constructed
in the terms of initial and final values of the parameters. Usual
finite dimensional integrable models correspond to the case when
the initial and final parameters coincide for all the operators
involved. It means the consideration of the trivial classical
dynamics. Algebraically, the conditions of the trivialization are
the origin of e.g. Baxter's curve for the chiral Potts model
\cite{CPM,BS-cpm}, or of the spherical triangle parameterization
for the Zamolodchikov-Bazhanov-Baxter model \cite{mss-vertex}.

One may hope to achieve
the progress in the spin integrable models by including into the consideration
a \emph{nontrivial}
classical dynamics. This dynamics will complicate the
situation, but sometimes more complicated
systems in the mathematics and physics may be solved more easily
-- proverbial ``a problem should be complicated until it becomes
trivial'' \cite{strogavnov}.

The quantum relativistic Toda chain at the root of unity is one of
the simplest examples of such combined classical/spin system. This
model was investigated in the preprints \cite{sp-1,sp-2,sp-3}. The
present paper is an attempt to present the detailed and clarified
exposition of this investigation.

We will use the language of the quantum inverse scattering method.
Besides the formulation of the model in the terms of $L$-operators
and transfer matrices, it implies the quantum intertwining
relation, allowing one to construct Baxter's  $Q$-operator.
Eigenvalues of the transfer matrix and $Q$-operator satisfy the Baxter, or
$TQ=Q'+Q''$, equation. Baxter equation may hardly be
solved explicitly for arbitrary $N$ and for finite length of the
chain $M$. It is the key relation for the investigation of the model in
the thermodynamic limit.  We will not investigate the
Baxter equation in this paper even in the thermodynamic limit, we will restrict
ourselves by a derivation of it.

Since any quantum Toda-type chain is the realm of the functional
Bethe ansatz, the problem of eigenstates of the transfer matrix is
related to the problem of the eigenstates of off-diagonal element
of the monodromy matrix, see e.g.
\cite{Sklyanin-sep,Sklyanin-fun,Lebedev} for the details. This is
true in our case as well. It is also known that Baxter's $Q$-operators
in the classical
limit are related to the B\"acklund
transformations of the classical chains, see e.g.
\cite{Sklyanin-rev,KS-manybody,kss-dst}.

In the following paragraph we describe in the shortest way the
subject of the present paper.

The main feature of the quantum relativistic Toda chain at the
root of unity with the nontrivial classical counterpart is that
the quantum intertwining relation contains the Darboux
transformation (see \cite{Sklyanin-rev}) for the $N$th powers of
the Weyl elements. So the B\"acklund transformation is the
classical counterpart of the finite dimensional $Q$-operator at
the root of unity automatically. Such $Q$-operators we call the
modified $Q$-operators, because they do not form the commutative
family. Usual $Q$-operators are their particular cases. Modified
$Q$-operators solve in general the isospectrality problem for the
spin chain. Starting from a homogeneous quantum chain, one may
create some number of solitons in parameterization of the
inhomogeneities using the B\"acklund transformations. The finite
dimensional counterpart of this procedure is a product of the
modified $Q$-operators, giving a similarity transformation of the
initial homogeneous transfer matrix. The term ``soliton'' is used
here in Hirota's sense: the system of the $N$th powers of Weyl's
elements may be parameterized by $\tau$-functions, obeying a
system of discrete equations. A solitonic solution of this system
is given by the Hirota-type expressions. Solitonic
$\tau$-functions contain extra free complex parameters -- the
amplitudes of solitonic partial waves. Generally speaking, these
parameters are a kind of ``times'' of the classical model
conjugated to the commutative set of classical hamiltonians. In
the other interpretation the amplitudes stand for a point on a
jacobian of a classical spectral curve in its rational limit.
Besides the complete solution of the quantum isospectrality
problem, these degrees of freedom appeared to be useful in an
unusual  sense. The amplitudes may be chosen so that one component
of a classical Backer-Akhiezer function becomes zero -- this is
related to the classical separation of variables. The
corresponding finite dimensional similarity operator becomes a
projector to an eigenvector of off-diagonal element of the
monodromy matrix, i.e. the base of the quantum separation of
variables. Moreover, since the similarity operator is the product
of $Q$-operators, Sklyanin's formula with the product of
\emph{eigenfunctions} $Q$ appears in the direct way.

This paper is devoted to: formulation of the combined
classical/spin model -- the quantum relativistic Toda chain at the
$N$th root of unity; its classical Darboux transformation/spin
intertwining relation; derivation and parameterization of
B\"acklund transformation/modified $Q$-operator; derivation of the
Baxter equation; explicit parameterization of the quantum
separating operator.

\section{Formulation of the model}

In this section we formulate the model called the
quantum relativistic Toda chain at the root of unity.

\subsection{$L$-operators}

Let the chain be formed by $M$ sites with the periodical boundary
conditions. $m$-th site of the Toda chain is described by the
following local $L$-operator:
\begin{equation}\label{l-toda}
\ds\ltoda_m(\lambda)\;\stackrel{def}{=}\;
\left(\begin{array}{rcr} \ds
1\,+\,{\kappa\over\lambda}\,\uop_m^{}\,\wop_m^{} &,&
\ds -\,{\omeg\over\lambda}\,\uop_m^{} \\
&&\\
\ds \wop_m^{} &,& 0 \end{array}\right)\;,
\end{equation}
where $\lambda$ is the spectral parameter  and $\kappa$ is an
extra complex parameter, common for all sites, i.e. a modulus of
eigenstates. Elements $\uop_m$ and $\wop_m$ form the ultra-local
Weyl algebra,
\begin{equation}\label{Weyl-algebra}
\ds \uop_m^{}\cdot\wop_m^{}\,=\,
\omega\;\wop_m^{}\cdot\uop_m^{}\;,
\end{equation}
and $\uop_m$, $\wop_m$ for different sites commute. Weyl's
parameter $\omega$ is the primitive root of unity,
\begin{equation}
\ds\omega\;=\;\EXP^{2\,\pi\,i/ N}\;,\;\;\;
\omeg\;=\;\EXP^{i\,\pi/N}\;.
\end{equation}
$N$ is an arbitrary positive integer greater then one and
common for all sites. The $N$th powers of the Weyl elements are
centers of the algebra. We will deal with the finite dimensional
unitary representation of the Weyl algebra
\begin{equation}\label{uw-XZ}
\ds \uop_m\;=\;u_m\,\xop_m\;,\;\;
\wop_m\;=\;w_m\,\zop_m\;,
\end{equation}
where $u_m$ and $w_m$ are $\mathbb{C}$-numbers, and
\begin{equation}\label{tensor}
\ds\xop_m = 1\otimes  ... \otimes
\underbrace{\xop}_{m-\textrm{th}\atop\textrm{place}}
\otimes ...\otimes 1\,,
\;\;\;
\zop_m = 1\otimes ... \otimes
\underbrace{\zop}_{m-\textrm{th}\atop\textrm{place}}
\otimes ... \otimes 1\;.
\end{equation}
A convenient representation of $\xop$ and $\zop$ in the
$N$-dimensional vector space $|\alpha\rangle=|\alpha\mod N\rangle$
is
\begin{equation}\label{representation}
\ds \xop\;|\alpha\rangle\;=\;|\alpha\rangle\;\omega^\alpha
\;,\;\;\;
\zop\;|\alpha\rangle\;=\;|\alpha+1\,\rangle\;,\;\;\;
\langle\alpha|\beta\rangle\;=\;\delta_{\alpha,\beta}\;.
\end{equation}
Thus $\xop$ and $\zop$ are $N\times N$ dimensional matrices,
normalized to the unity ($\xop^N=\zop^N=1$), and the
$N$th powers of the local Weyl elements are $\mathbb{C}$-numbers
\begin{equation}\label{uw-N-powers}
\ds \uop_m^N\;=\;u_m^N\;,\;\;\; \wop_m^N\;=\;w_m^N\;.
\end{equation}
In general, all $u_m$ and $w_m$ are different, such the chain is
the inhomogeneous one.

Variables $u_m^N$ and $w_m^N$ form the classical counterpart of
the quantum relativistic Toda chain. Define the classical $L$-operator
as
\begin{equation}\label{L-toda}
\ds\Ltoda_m(\lambda^N)\,\stackrel{def}{=}\,
\left(\begin{array}{rcr} \ds
1\,+\,{\kappa^N\over\lambda^N}\,u_m^N w_m^N &, & \ds
{u_m^N\over\lambda^N}\\
&&\\
\ds w_m^N &,& 0
\end{array}\right)\;.
\end{equation}

\subsection{Transfer matrices and integrability}

Ordered product of the quantum $L$-operators
\begin{equation}\label{l-monodromy}
\ds\montoda(\lambda)\;\stackrel{def}{=}\;
\ltoda_1(\lambda)\;\ltoda_2(\lambda)\;\cdots\;
\ltoda_M(\lambda)\;=\;
\left(\begin{array}{cc} \ds
\mathbf{a}(\lambda) & \ds
\mathbf{b}(\lambda)\\
&
\\ \ds \mathbf{c}(\lambda) &
\mathbf{d}(\lambda)\end{array}\right)
\end{equation}
and its trace
\begin{equation}\label{transfer}
\ds \tp(\lambda)\;=\;\mathbf{a}(\lambda)+\mathbf{d}(\lambda) \;=\;
\sum_{k=0}^{M}\; \lambda^{-k}\;\tp_k
\end{equation}
are the monodromy and the transfer matrices of the quantum model.

The integrability of the quantum chain is provided by the
intertwining relation in the auxiliary two-dimensional spaces
\begin{equation}\label{RLL}
\ds R(\lambda,\mu)\;\ltoda(\lambda)\otimes\ltoda(\mu) \;=\;
(1\otimes \ltoda(\mu))\;(\ltoda(\lambda)\otimes 1)\;R(\lambda,\mu)\;.
\end{equation}
The tensor product of two $2\times 2$
matrices with the identical Weyl algebra entries
$\uop,\wop,\kappa$ but different spectral parameters $\lambda$
and $\mu$ is implied in (\ref{RLL}), and
\begin{equation}\label{six-vertex}
\ds R(\lambda,\mu)\;=\;\left(\begin{array}{cccc} \ds
\lambda\,-\,\omega\,\mu & 0 & 0 & 0 \\
0 & \ds \lambda\,-\,\mu & \ds \mu\,(1\,-\,\omega) & 0 \\
0 & \ds \lambda\,(1\,-\,\omega) & \ds
\omega\,(\lambda\,-\,\mu) & 0 \\
0 & 0 & 0 & \ds \lambda\,-\,\omega\,\mu
\end{array}\right)
\end{equation}
is the slightly twisted six-vertex trigonometric $R$-matrix, used
e.g. in \cite{Tarasov-cyclic} . Equation (\ref{RLL}) may be
verified directly. The repeated use of the intertwining relation
(\ref{RLL}) gives the analogous relation for the monodromy
matrices (\ref{l-monodromy}), leading to the commutativity of the
transfer matrices (\ref{transfer}) with different spectral
parameters but with the identical elements $\uop_m,\wop_m$ and
$\kappa$
\begin{equation}\label{comm}
\ds \Bigl[\;\tp(\lambda,\kappa\,;\{u_m,w_m\})\;,\;
\tp(\mu,\kappa\,;\{u_m,w_m\})\;\Bigr]\;=\;0\;,
\end{equation}
where we emphasize the dependence of
$\tp(\lambda)=\tp(\lambda,\kappa\,;\{u_m,w_m\}_{m=1}^M)$ on  the
given set of $\mathbb{C}$-valued parameters $u_m,w_m$, $m=1...M$.

In the spectral decomposition of the transfer matrix
(\ref{transfer}) the utmost operators are
\begin{equation}\label{utmost}
\ds \tp_0=1\;,\;\;\;
\tp_M=\kappa^M\,\prod_{m=1}^M\,(-\omega^{1/2}u_mw_m)\,
\mathbf{Y}\;,
\end{equation}
where
\begin{equation}\label{Y}
\ds\mathbf{Y} =
\prod_{m=1}^M\,(-\omega^{-1/2}\xop_m\zop_m)\;,\;\;\;
\mathbf{Y}^N=1\;.
\end{equation}

Transfer matrix (\ref{transfer}) is invariant with respect to a
gauge transformation of $L$-operators (\ref{l-toda}). Let $G$ be $2\times 2$
matrix with $\mathbb{C}$-valued coefficients, then the
transformation
\begin{equation}\label{gauge-transform}
\ds \ell_m\;\mapsto\; G\,\ell_m\,G^{-1}\;\Rightarrow\;
\montoda\;\mapsto\;G\,\montoda\,G^{-1}
\end{equation}
does not change the transfer matrix. For example,
\begin{equation}
\ds G\;=\;\left(\begin{array}{cc} \ds 1 & \ds 0 \\ \ds 0 &
g\end{array}\right)
\end{equation}
produces the transformation of the parameters
\begin{equation}\label{gauge-invariance}
\ds u_m\;\mapsto\;g^{-1}u_m\;,\;\;\;w_m\;\mapsto\;g\,w_m\;.
\end{equation}
Taking into account this freedom and also the possibility of
redefinition of the spectral parameter $\lambda$, entering
$L$-operator (\ref{l-toda}) always as $\lambda^{-1}u_m$, we may
impose,  without loss of generality, the following conditions for
the parameters $\{u_m,w_m\}_{m=1}^M$
\begin{equation}\label{gauge-fix}
\ds \prod_{m=1}^M\,(-\omega^{1/2}u_m) \;=\; \prod_{m=1}^M(-w_m)
\;=\;1 \;.
\end{equation}

\subsection{Classical monodromy matrix}

Define the monodromy of the classical Lax matrices (\ref{L-toda})
analogously to (\ref{l-monodromy}):
\begin{equation}\label{Mon-toda}
\ds\Montoda\;\stackrel{def}{=}
\;\Ltoda_1\;\Ltoda_2\;\cdots\;\Ltoda_M\;=\;
\left(\begin{array}{cc} \ds A(\lambda^N) & \ds
B(\lambda^N)\\
&\\
\ds C(\lambda^N) & \ds D(\lambda^N)
\end{array}\right)\;.
\end{equation}
For our purposes we have to mention the relation
between the elements of the quantum monodromy matrix
(\ref{l-monodromy}) and the classical ones. Note at first,
relation (\ref{RLL}) provides the commutativity
\begin{equation}
\ds \left[\mathbf{a}(\lambda),\mathbf{a}(\mu)\right]\,=\,
\left[\mathbf{b}(\lambda),\mathbf{b}(\mu)\right]\,=\,
\left[\mathbf{c}(\lambda),\mathbf{c}(\mu)\right]\,=\,
\left[\mathbf{d}(\lambda),\mathbf{d}(\mu)\right]\,=\,0\;.
\end{equation}
The spectra of $\mathbf{a}(\lambda)$, $\mathbf{b}(\lambda)$,
$\mathbf{c}(\lambda)$ and $\mathbf{d}(\lambda)$ may be calculated
by means of the following
\begin{prop}\label{prop-abcd}
\begin{equation}
\begin{array}{l}
\ds \prod_{n\in\mathbb{Z}_N}\;\mathbf{a}(\omega^n\lambda)
\;=\;A(\lambda^N)\;,\;\;\;
\prod_{n\in\mathbb{Z}_N}\;\mathbf{b}(\omega^n\lambda) \;=\;B(\lambda^N)\;,\\
\\
\ds \prod_{n\in\mathbb{Z}_N}\;\mathbf{c}(\omega^n\lambda)
\;=\;C(\lambda^N)\;,\;\;\;
\prod_{n\in\mathbb{Z}_N}\;\mathbf{d}(\omega^n\lambda)
\;=\;D(\lambda^N)\;.
\end{array}
\end{equation}
\end{prop}
Theorems of such kind were proved in \cite{Tarasov-cyclic} for the
general case of cyclic representations like (\ref{representation})
of the quantum affine algebras. An alternative proof follows from
the results of ref. \cite{Sergeev}.

Let us consider the homogeneous chain with $u_m=u$ and $w_m=w$.
Due to (\ref{gauge-invariance},\ref{gauge-fix}), we may fix the
parameters of the homogeneous chain as
\begin{equation}\label{uw-homo}
\ds u_m^{}\;=\;-\,\omega^{-1/2}\;\;\;\textrm{and}\;\;\;
w^{}_m\;=\;-\,1\;.
\end{equation}
Then
\begin{equation}
\ds\Montoda\;=\;L(\lambda^N)^M\;,\;\;\;
L(\lambda^N)\;=\;\left(\begin{array}{rr}
\ds 1-{\kappa^N\over\lambda^N} & \ds -{\epsilon\over\lambda^N}\\
\ds \epsilon & 0\end{array}\right)\;,
\end{equation}
where $\epsilon=(-1)^N$, and the matrix elements of $\Montoda$
may be calculated with the help of projector decomposition of
$L(\lambda^N)$.  Namely, let $x_1$ and $x_2$ are two eigenvalues
of $L(\lambda^N)$:
\begin{equation}\label{x1x2}
\ds x_1x_2\;=\;{1\over\lambda^N}\;,\;\;\;
x_1+x_2\;=\;1-{\kappa^N\over\lambda^N}\;.
\end{equation}
Then $L(\lambda^N)$ may be presented in the form
$L(\lambda^N)=D\,\textrm{diag}(x_1,x_2)\, D^{-1}$, therefore
$\Montoda=D\,\textrm{diag}(x_1^M,x_2^M)\, D^{-1}$, and the final
answer reads
\begin{equation}\label{Tx1x2}
\ds\Montoda\;=\;\left(\begin{array}{rcr} \ds
{x_1^{M+1}-x_2^{M+1}\over x_1^{}-x_2^{}} &,& \ds
-\epsilon x_1^{}x_2^{}\,{x_1^M-x_2^M\over x_1^{}-x_2^{}}\\
&&\\ \ds \epsilon\,{x_1^M-x_2^M\over x_1^{}-x_2^{}} &,& \ds
{x_1^{}x_2^{M}-x_2^{}x_1^{M}\over
x_1^{}-x_2^{}}\end{array}\right)\;.
\end{equation}
Now one may calculate $\ds {x_1^M-x_2^M\over x_1^{}-x_2^{}}$
explicitly as a function of $\lambda^N$. First of all, it is an
$M-1$-th power polynomial with respect to $\ds{1\over\lambda^N}$,
so
\begin{equation}\label{x1x2lambda}
\ds {x_1^M-x_2^M\over x_1^{}-x_2^{}}\;=\; C_0\;
\prod_{k=1}^{M-1}\;
\left(1\;-\;{\lambda_k^N\over\lambda^N}\right)\;.
\end{equation}
In the limit when $1/\lambda^N\mapsto 0$, one has $x_1=1$ and
$x_2=0$, see (\ref{x1x2}). Therefore $C_0=1$. The roots
$\lambda^N=\lambda_k^N$ of (\ref{x1x2lambda}) correspond to the
case when
\begin{equation}
\ds x_1^M=x_2^M\;,\;\;\;x_1^{}\neq x_2^{}\;\;\; \Rightarrow\;\;\;
{x_2\over x_1}\;=\;\EXP^{2\,i\,\phi_k}\;,
\end{equation}
where $\phi_k$, $k=1...M-1$, belong to the following set $\Set$:
\begin{equation}\label{phi-k}
\ds\phi_k\;\in\;\Set\;\stackrel{def}{=}\; \{{\,\pi\over
M}\,,\;{2\pi\over M}\,,\;...\;,\;{(M-1)\pi\over M}\,\}\;.
\end{equation}
The ordering of $\phi_k\in\Set$ is not essential. Introduce now
three useful functions $\Om_\phi^{}$, $\Om_\phi^*$ and
$\Lambda_\phi$:
\begin{equation}\label{phi-parametrization}
\ds\left\{\begin{array}{l} \ds
\Om_\phi^{}\;=\;\EXP^{i\,\phi}\,\left(
\sqrt{\cos^2\phi\,+\,\kappa^N}\;+\;\cos\phi\right)\;,\\
\\
\Om^*_\phi\;=\;\EXP^{-i\,\phi}\,\left(
\sqrt{\cos^2\phi\,+\,\kappa^N}\;+\;\cos\phi\right)\;,\\
\\
\ds \Lambda_\phi\;=\;\Om^{}_\phi\;\Om^*_\phi\;.
\end{array}\right.
\end{equation}
Expressions for $\Delta$ and $\Delta^*$ uniformize the rational
curve
\begin{equation}\label{Delta-curve}
\ds\Delta\Delta^*\;=\;\Delta\,+\,\Delta^*\,+\,\kappa^N\;\;
\textrm{in the terms of}\;\; \ds
{\Delta\over\Delta^*}\;=\;\EXP^{2i\phi}\;.
\end{equation}
Comparing (\ref{x1x2}) and (\ref{Delta-curve}), we conclude
\begin{equation}
\ds \lambda_k^N\;=\;\Lambda_{\phi_k}\;,\;\;\;\phi_k\in\Set\;.
\end{equation}
The application of Proposition \ref{prop-abcd} on page \pageref{prop-abcd}
to the the homogeneous chain gives in particular
\begin{equation}\label{b-spectrum}
\ds\prod_{n\in\mathbb{Z}_N}
\;\mathbf{b}(\omega^n\lambda)\;=\;-{\epsilon\over\lambda^N}\,
\prod_{k=1}^{M-1}\;
\left(1\,-\,{\Lambda_{\phi_k}\over\lambda^N}\right)\;,\;\;\;\phi_k\in\Set\;.
\end{equation}

\subsection{Eigenvalues and eigenstates}

The natural aim of the investigation of any integrable quantum
chain is to calculate the spectrum of the transfer matrix and to
construct its eigenstates at least for the homogeneous chain. In this
subsection we will fix several notations necessary for the
spectral problems and for the functional Bethe ansatz.

For given transfer matrix $\tp(\lambda)$ let $t(\lambda)$ be its
eigenvalue for the right eigenvector $|\Psi_t\rangle$ and for the
left eigenvector $\langle\Psi_t|$:
\begin{equation}\label{t-eigen}
\ds \tp(\lambda)\;|\Psi_t\rangle\;=\;|\Psi_t\rangle\;t(\lambda)\;,
\;\;\;
\langle\Psi_t|\;\tp(\lambda)\;=\;t(\lambda)\;\langle\Psi_t|\;.
\end{equation}
Here $t(\lambda)=\sum_{k=0}^M\,\lambda^{-k}\,t_k$,
$t_k$ are eigenvalues of the complete commutative set of
$\tp_k$  (\ref{transfer}). Due to
(\ref{utmost},\ref{Y}), $t_0\equiv 1$ and
$t_M=(-\kappa)^M\,\omega^\gamma$, where $\omega^\gamma$ is the
eigenvalue of $\mathbf{Y}$. The subscript `$t$' in the notations
of the eigenvectors stands for the $M$-components vector
$\{t_k\}$, $k=1,...,M$. The eigenvectors do not depend
on $\lambda$, but depend on $\kappa$ and $u_m,w_m$ for the
inhomogeneous chain. In general $\tp_k$ are not hermitian, so left
and right eigenvectors are not conjugated. Nevertheless, let us
imply that $|\Psi_t\rangle$ and $\langle\Psi_t|$ are dual
complete basises,
\begin{equation}
\ds \tp(\lambda) \;=\; \sum_t\;
|\Psi_t\rangle\;t(\lambda)\;\langle\Psi_t|
\;\;\;\textrm{and}\;\;\;
1\;=\;\sum_t\;|\Psi_t\rangle\;\langle\Psi_t|\;,
\end{equation}
where the summations over $t$ is by definition the summation over
all $N^M$ possible sets of the eigenvalues $\{t_k\}_{k=1}^M$.

All the observables are defined in the local basis initially
(\ref{tensor}):
\begin{equation}\label{local-basis}
\ds |\alpha\rangle\;=\;|\alpha_1\rangle\otimes|\alpha_2\rangle
\otimes\cdots\otimes|\alpha_M\rangle\;.
\end{equation}
The dual basis is defined via
\begin{equation}
\ds
\langle\alpha|\beta\rangle\;=\;\prod_{m=1}^M\,\delta_{\alpha_m,\beta_m}\;,\;\;\;
1\;=\;\sum_{\{\alpha_m\}}\,|\alpha\rangle\,\langle\alpha|\;.
\end{equation}
Calculation of the eigenvector $|\Psi_t\rangle$ means the
calculation of the matrix elements $\langle\alpha|\Psi_t\rangle$
\begin{equation}
\ds |\Psi_t\rangle \;=\; \sum_{\{\alpha_m\}}\,|\alpha\rangle\,
\langle\alpha|\Psi_t\rangle\;.
\end{equation}

Let us formulate now the isospectrality problem. Let
$\{u^{}_m,w^{}_m\}_{m=1}^M$ and $\{u_m',w_m'\}_{m=1}^M$ be two
sets of parameters of Weyl elements for the quantum relativistic
Toda chain at the root of unity. These sets are isospectral if
\begin{equation}\label{isospectral}
\ds t(\lambda;\{u_m^{},w_m^{}\}_{m=1}^M) \;=\;
t(\lambda;\{u_m',w_m'\}_{m=1}^M)\;.
\end{equation}
Let $|\Psi_t^{}\rangle$ and $|\Psi'_t\rangle$ be the eigenvectors of
two isospectral transfer matrices $\tp(\lambda)$ and
$\tp'(\lambda)$ with the same eigenvalue $t(\lambda)$. Two isospectral
transfer matrices must be related by  a similarity operator
$\mathbf{K}$
\begin{equation}\label{K-similarity}
\ds \tp(\lambda;\{u_m^{},w_m^{}\}_{m=1}^M)\;\mathbf{K}\;=\;
\mathbf{K}\;\tp(\lambda;\{u_m',w_m'\}_{m=1}^M)\;,
\end{equation}
such that
\begin{equation}\label{K-decomposition}
\ds \mathbf{K}\;=\;\sum_{t}\; |\Psi_t^{}\rangle\; K_t\;
\langle\Psi'_t|\;.
\end{equation}
Here $K_t$ are arbitrary nonzero values.  We will prove further
that two transfer matrices are isospectral iff the traces of their
classical monodromy matrices coincide.

The eigenvalues of the transfer matrix obey several functional
relations. Sometimes they may be solved in the thermodynamic limit
or at least some physical properties of the chain may be derived
from the functional equations in the thermodynamic limit.

The eigenstates for quantum Toda chain, relativistic as well as
usual, and in general the eigenstates for models based on
the local Weyl algebra at root of the unity may be constructed with
the help of so-called functional Bethe ansatz, or quantum
separation of variables \cite{Sklyanin-sep,Sklyanin-fun,Lebedev}.
Eigenstates of off-diagonal element of the monodromy matrix
$\mathbf{b}(\lambda)$ (\ref{l-monodromy}) play the important
r\^ole in the method of the functional Bethe ansatz. It is useful
to parameterize the spectrum of $\mathbf{b}(\lambda)$ by its zeros
$\lambda_k$, $k=1...M-1$.  It follows from (\ref{b-spectrum}) for
the homogeneous chain
\begin{equation}
\ds\lambda_k^N\;=\;\Lambda_{\phi_k}\;,\;\;\;k=1...M-1,\;\;\;\phi_k\in\Set.
\end{equation}
The monodromy matrix (\ref{l-monodromy}) has the
structure
\begin{equation}
\ds\montoda(\lambda)\;=\;\montoda'(\lambda)\,\cdot\,\ell_M(\lambda)\;,
\end{equation}
where $\montoda'(\lambda)$ is the monodromy of the first $M-1$
sites. Then
\begin{equation}
\ds \mathbf{b}(\lambda) \;=\; \mathbf{a}'(\lambda)\,
\left(-\,{\omega^{1/2}\over\lambda}\,\uop_M\right)\;.
\end{equation}
It is more convenient to work with the eigenvectors of the operator
\begin{equation}\label{btilde}
\ds \widetilde{\mathbf{b}}(\lambda)\;\stackrel{def}{=}\;
\mathbf{b}(\lambda)\,\wop_M\;= \;\mathbf{a}'(\lambda)\,
\left(-\,{\omega^{1/2}\over\lambda}\,\uop_M\,\wop_M\right)\;.
\end{equation}
Note that operator $\mathbf{Y}$ (\ref{Y}) commute with
$\widetilde{\mathbf{b}}$. We denote the eigenvectors of $\widetilde{\mathbf{b}}$
as $\ds |\{\lambda_k\}_{k=1}^{M-1},\gamma\rangle$ and
define them by the following relation
\begin{equation}\label{b-eigen}
\ds
\mathbf{b}(\lambda_j)\,|\{\lambda_k\}_{k=1}^{M-1},\gamma\rangle
\,=\,0\;,\;\;\;\lambda_j\in\{\lambda_k\}_{k=1}^{M-1}\;,
\end{equation}
and
\begin{equation}
\ds \mathbf{Y}\,|\{\lambda_k\}_{k=1}^{M-1},\gamma\rangle\,=\,
|\{\lambda_k\}_{k=1}^{M-1},\gamma\rangle\,\omega^\gamma\;.
\end{equation}

The functional Bethe ansatz method, being applied to
the quantum relativistic Toda chain at root of unity, implies the
following structure of the matrix element (see
\cite{Sklyanin-fun}):
\begin{equation}\label{q-sep}
\ds \langle\Psi_t|\{\lambda_k\}_{k=1}^{M-1},\gamma\rangle\;=\;
\textrm{const}\;\prod_{k=1}^{M-1}\;q_t(\lambda_k)\;.
\end{equation}
$q_t(\lambda)$ is a function, which depends on the spectrum of
$\tp(\lambda)$, and obeys a functional equation, called the Baxter
$T-Q$ equation. Operator $\mathbf{Y}$, being the member of $\tp$
(\ref{utmost}), has the eigenvalue $\omega^\gamma$ in all the
components of formula (\ref{q-sep}). The explicit form of Baxter's
equation and the definition of $q_t(\lambda)$ will be given later.

One obtains usually formula (\ref{q-sep}) from equation
(\ref{RLL}), see \cite{Sklyanin-sep,Sklyanin-fun,Lebedev}. In this
paper we will obtain (\ref{q-sep}) in a different way: we will
construct first so-called modified $\Q$-operator, whose spectrum
is related to $q_t$. Then a projector to
$|\{\lambda_k\}_{k=1}^{M-1},\gamma\rangle$ will be obtained using
these modified $\Q$-operators.

\section{Intertwining relations}

The quantum relativistic Toda chain $L$-operator (\ref{l-toda}) is
$2\times 2$ matrix, whose matrix elements are
$N\times N$ matrices. According to the conventional terminology,
the two-dimensional vector space is called the auxiliary space,
while $N$-dimensional one is called the quantum space. Equation
(\ref{RLL}) is the intertwining relation in the auxiliary spaces.
In this section we will investigate the other type of the
intertwining relations -- in the quantum spaces.

It is easy to check that two $L$-operators (\ref{l-toda}) can not be
intertwined in the quantum spaces (i.e. their intertwiner is exactly zero).
In the case of the usual quantum Toda chain one needs an extra auxiliary
$L$-operator (in its simplest form it is known as the $L$-operator for the
``dimer self-trapping model'' \cite{Sklyanin-rev}). The same situation holds
in our case: we need an auxiliary $L$-operator that may be intertwined with
$L$-operator (\ref{l-toda}) in the quantum spaces. In general, this
auxiliary $L$-operator is Bazhanov-Stroganov's $L$-operator for the
six-vertex model at the root of unity \cite{BS-cpm}. This fact is the origin
of a relationship between the relativistic Toda chain at the $N$th root of
unity and the $N$-state chiral Potts model.
The Bazhanov-Stroganov $L$-operator may be simplified
to a ``relativistic dimer self-trapping'' $L$-operator.

In this section we introduce this auxiliary $L$-operator and write
down the quantum intertwining relation. Being written in an
appropriate way, this will give us a key for all further
investigations.

\subsection{Auxiliary $L$-operator}

Define the auxiliary quantum $L$-operator as follows:
\begin{equation}\label{ldst-uwkappa}
\ds
\ldst^{}_\ff(\lambda,\lambda_\ff)\;\stackrel{def}{=}\;
\left(\begin{array}{rcr} \ds
1\,-\,\omeg\kappa_\ff^{}{\lambda_\ff^{}\over\lambda} \wop_\ff^{}
&,& \ds
-{\omeg\over\lambda}(1-\omeg\kappa_\ff^{}\wop_\ff^{})\uop_\ff^{}
\\
&&\\
\ds -\omeg\lambda_\ff^{}\uop_\ff^{-1}\wop_\ff^{} &,& \ds
\wop_\ff^{}
\end{array}\right)\;.
\end{equation}
Here $\lambda$ and $\lambda_\ff$ are two spectral parameters
(actually, up to a gauge transformation (\ref{gauge-transform}),
$\ldst$ depends on their ratio). $\kappa_\ff$ is a module
analogous to $\kappa$. $\uop_\ff$ and $\wop_\ff$ are additional Weyl elements,
defined analogously to
(\ref{Weyl-algebra},\ref{uw-XZ},\ref{tensor},\ref{representation})
\begin{equation}\label{uw-XZ-phi}
\ds\uop_{\ff}\;=\;u_\ff\;\xop_\ff\;,\;\;\;
\wop_\ff\;=\;w_\ff\;\zop_\ff\;.
\end{equation}
In these notations the subscript $\ff$ labels
the additional Weyl pair in the tensor product
(\ref{tensor}).

Classical counterpart of $\ldst$ is by definition
\begin{equation}\label{L-dst}
\ds \Ldst_\ff(\lambda^N,\lambda_\ff^N)
\;\stackrel{def}{=}\;\left(\begin{array}{rcr} \ds
1+\kappa_\ff^{N}{\lambda_\ff^N\over\lambda^N}w_\ff^{N} &,&
\ds{u_\ff^N\over\lambda^N}(1+\kappa_\ff^N w_\ff^N)\\
&&\\
\ds \lambda_\ff^N{w_\ff^N\over u_\ff^N} &,& w_\ff^N
\end{array}\right)\;.
\end{equation}

$L$-operators (\ref{ldst-uwkappa}) are intertwined in their two
dimensional auxiliary vector spaces by the same six-vertex
trigonometric $R$-matrix (\ref{six-vertex}).
There exists also the fundamental quantum
intertwiner for (\ref{ldst-uwkappa}). It is the $R$-matrix for the
chiral Potts model such that two rapidities are fixed to special
singular values. The details are useless in this paper. But the
method described here may be applied directly to the
model, defined by (\ref{ldst-uwkappa}). This is the subject of
forthcoming papers.

\subsection{Quantum intertwining relation}

We are going to write out some quantum intertwining relation for
$L$-operators (\ref{l-toda}) as well as for the whole monodromy
matrix (\ref{l-monodromy}). So we use notations, applicable for
the recursion in $m$. Also in this section we will point out
parameters $u_m,w_m$ and $u_\ff,w_\ff$ as the arguments of $\ell_m$ and
$\ldst_\ff$.

\begin{prop}\label{prop-intertwining}
There exists unique (up to a constant multiplier) $N^2\times N^2$
matrix $\R_{m,\ff}(\lambda_\ff)$, such that $\R_{m,\ff}$,
$\ltoda_m$ and $\ldst$ obey the modified intertwining relation
\begin{equation}\label{op-recursion}
\ds\begin{array}{l}
\ds\ldst_\ff^{}(\lambda,\lambda_\ff;u_{\ff,m}^{},w_{\ff,m}^{})\cdot
\ltoda_m^{}(\lambda;u_m^{},w_m^{})\;\R_{m,\ff}^{}(\lambda_\ff)\;=\\
\\
\ds =\;
\R_{m,\ff}^{}(\lambda_\ff)\;\ltoda_m^{}(\lambda;u_m',w_m')\cdot
\ldst_\ff^{}(\lambda,\lambda_\ff;u_{\ff,m+1}^{},w_{\ff,m+1}^{})\;,
\end{array}
\end{equation}
if and only if their classical counterparts of $\Ltoda_m$ and
$\Ldst_\phi$ obey the following Darboux relation:
\begin{equation}\label{LL-darboux}
\ds\Ldst_\phi(\lambda^N,\lambda_\ff^N;u_{\ff,m}^{N},w_{\ff,m}^{N})
\Ltoda_m(\lambda^N;u_m^{N},w_m^{N})\;=\;
\Ltoda_m(\lambda^N;u_m^{\prime N},w_m^{\prime
N}) \Ldst_\phi(\lambda^N,\lambda_\ff^N;u_{\ff,m+1}^{N},w_{\ff,m+1}^{N})\;,
\end{equation}
valid in each order of $\lambda^N$.
\end{prop}
\noindent\emph{Proof}.  Consider the following equation
\begin{equation}\label{fundamental}
\ds \ldst_\ff^{}(\lambda;\uop_\ff^{},\wop_\ff^{})\cdot
\ltoda_m^{}(\lambda;\uop_m^{},\wop_m^{})\;=\;
\ltoda_m^{}(\lambda;\uop_m',\wop_m')\cdot
\ldst_\ff^{}(\lambda;\uop_\ff',\wop_\ff')
\end{equation}
and equate its coefficients in each order of $\lambda$. In
(\ref{fundamental} we used the Weyl elements as the formal
arguments of $L$-operators (\ref{l-toda}) and
(\ref{ldst-uwkappa}). This system of equations has the unique
solution with respect to the ``primed'' operators:
\begin{equation}\label{the-mapping}
\ds\left\{\begin{array}{l}
\ds\uop_m'\;=\;{\kappa_\ff\over\kappa}\,\uop_\ff^{}\;,\\
\\
\ds\wop_m'\;=\;
\wop_m^{}\wop_\ff^{}\,-\,\omeg\lambda_\ff^{}\uop_\ff^{-1}
\wop_\ff^{}\;,\\
\\
\ds\uop_\ff'\;=\; \left(\lambda_\ff\,+\,\kappa\uop_m\wop_m\,-\,
\omeg\wop_m\uop_\ff\right)^{-1}\lambda_\ff^{}\uop_m^{}\;,\\
\\
\ds\wop_\ff'\;=\; {\kappa\over\kappa_\ff}\,\uop_m\wop_m\,
\left(\wop_m^{}\uop_\ff^{}\,-\,\omeg\lambda_\ff^{}\right)^{-1}\;.
\end{array}\right.
\end{equation}
Obtaining (\ref{the-mapping}) from (\ref{fundamental}), we used
the exchange relations for $\uop_m,\wop_m,\uop_\ff,\wop_\ff$ without
assuming any exchange relations for ``primed'' operators.
One may check directly that the rational homomorphism
\begin{equation}\label{homomorphism}
\ds\uop_m^{},\wop_m^{},\uop_\ff^{},\wop_\ff^{}\;\mapsto\;
\uop_m',\wop_m',\uop_\ff',\wop_\ff'\;,
\end{equation}
given by (\ref{the-mapping}), is the automomorphism of the local
Weyl algebra.

Since the finite dimensional representations at root of unity are
considered (see eqs. (\ref{uw-XZ}) and (\ref{uw-XZ-phi})), we
may use the normalization of the operators entering
(\ref{fundamental}) as usual:
\begin{equation}\label{uwuw-ini}
\ds \uop_m^{}=u_m^{}\xop_m^{}\,,\; \wop_m^{}=w_m^{}\zop_m^{}\,,\;
\uop_\ff^{}=u_{\ff,m}^{}\xop_\ff^{}\,,\;
\wop_\ff^{}=w_{\ff,m}^{}\zop_\ff^{}\;,
\end{equation}
and
\begin{equation}\label{uwuw-fin}
\ds \uop_m'=u_m'\xop_m'\,,\; \wop_m'=w_m'\zop_m'\,,\;
\uop_\ff'=u_{\ff,m+1}^{}\xop_\ff'\,,\;
\wop_\ff'=w_{\ff,m+1}^{}\zop_\ff'\;,
\end{equation}
where all $\xop$, $\zop$ matrices are normalized to unity. With
this normalization, the $N$th powers of (\ref{the-mapping}) may
be calculated directly with the help of the following identity
\begin{equation}\label{fc-gen}
\ds (\uop\,+\,\wop)^N\;=\;\uop^N\,+\,\wop^N\;.
\end{equation}
The answer reads
\begin{equation}\label{ff-recursion}
\ds u_{\ff,m+1}^N\;=\;{\lambda_\ff^N\,u_m^N\over
\lambda_\ff^N+\kappa^N u_m^N w_m^N + w_m^N u_{\ff,m}^N}\;,\;\;\;
\ds w_{\ff,m+1}^N \;=\; {\kappa^N\over\kappa_\ff^N}\,
{u_m^Nw_m^N\over \lambda_\ff^N+w_m^N u_{\ff,m}^N}\;,
\end{equation}
and
\begin{equation}\label{uw-prime}
\ds u_m^{\prime N} \;=\;
{\kappa_\ff^N\over\kappa^N}\,u_{\ff,m}^{N}\;, \;\;\;\; w_m^{\prime
N} \;=\; {\kappa^N\over\kappa_\ff^N}\,{u_m^{N}w_m^{N}\over
u_{\ff,m}^N}\,{w_{\ff,m}^{N}\over w_{\ff,m+1}^{N}}\;.
\end{equation}
It is easy to check that (\ref{ff-recursion}) and
(\ref{uw-prime}) are the exact and the unique solution of
(\ref{LL-darboux}).

Automorphism (\ref{homomorphism}) with normalizations
(\ref{uwuw-ini}) and (\ref{uwuw-fin}) taken into account, gives
the automorphism
\begin{equation}
\ds\xop_m^{},\zop_m^{},\xop_\ff^{},\zop_\ff^{}\;\mapsto\;
\xop_m',\zop_m',\xop_\ff',\zop_\ff'\;,
\end{equation}
where, recall, all $\xop$, $\zop$ are normalized to unity finite
dimensional operators. This is provided by
(\ref{ff-recursion},\ref{uw-prime}) or by
(\ref{LL-darboux}). Therefore this automorphism must be the
internal one, i.e. due to the Shur lemma there must exist an
unique (up to a multiplier) $N^2\times N^2$ matrix $\R_{m,\ff}$:
\begin{equation}
\ds\xop_m' = \R_{m,\ff}^{}\xop_m^{}\R_{m,\ff}^{-1}\;,\;\;\;
\zop_m' = \R_{m,\ff}^{}\zop_m^{}\R_{m,\ff}^{-1}\;,
\end{equation}
and
\begin{equation}
\ds\xop_\ff' = \R_{m,\ff}^{}\xop_\ff^{}\R_{m,\ff}^{-1}\;,\;\;\;
\zop_\ff' = \R_{m,\ff}^{}\zop_\ff^{}\R_{m,\ff}^{-1}\;.
\end{equation}
The last two equations are equivalent to (\ref{op-recursion}).\hfill $\Box$

The local transformation
\begin{equation}\label{darboux-mapping}
\ds u_m^N,w_m^N,u_{\ff,m}^N,w_{\ff,m}^N\;\;\mapsto\;\; u_m^{\prime
N},w_m^{\prime N},u_{\ff,m+1}^N,w_{\ff,m+1}^N\;,
\end{equation}
given by eqs. (\ref{ff-recursion}) and (\ref{uw-prime}), is called
the Darboux transformation for the classical
relativistic Toda chain, see e.g. \cite{Sklyanin-rev} for the
analogous transformation for the usual Toda chain.
(\ref{ff-recursion}) and (\ref{uw-prime}) define the
mapping (\ref{darboux-mapping}) up to $N$th roots of unity. These
phases are the additional discrete parameters appeared when one
takes the $N$th roots in (\ref{darboux-mapping}). Note, the matrix
$\R_{m,\ff}$ is unique if all these roots are fixed.

In the next section we will give matrix elements of $\R_{m,\ff}$
in the basis (\ref{representation}).

\subsection{$\mathcal{Q}$-transformation and the isospectrality problem}

Relations (\ref{op-recursion}) and (\ref{LL-darboux}) may be
iterated for the whole chain. The functional counterpart, eq.
(\ref{LL-darboux}), gives
\begin{equation}\label{LT-darboux}
\ds\Ldst_\ff(u_{\ff,1}^{N},w_{\ff,1}^{N})
\Montoda(\{u_m^{N},w_m^{N}\}_{m=1}^M)\;=\; \Montoda(\{u_m^{\prime
N},w_m^{\prime N}\}_{m=1}^M)
\Ldst_\ff(u_{\ff,M+1}^{N},w_{\ff,M+1}^{N})\;,
\end{equation}
where $\Montoda$ is the classical monodromy matrix
(\ref{Mon-toda}) and the spectral parameters are implied. Values of
$u_m',w_m'$, $m=1,...,M$, and $u_{\ff,M+1},w_{\ff,M+1}$ in the
terms of $\lambda_\ff,\kappa_\ff$, $u_m^{},w_m^{}$ and
$u_{\ff,1},w_{\ff,1}$ must be obtained as the recursion
iterating  (\ref{ff-recursion},\ref{uw-prime}) for $m=1,...,M$.

For the periodic chain the cyclic boundary conditions for the
recursion have to be imposed,
\begin{equation}\label{cyclic-recursion}
\ds u_{\ff,M+1}\;=\;u_{\ff,1}\;,\;\;\;w_{\ff,M+1}\;=\;
w_{\ff,1}\;.
\end{equation}
Now suppose that (\ref{ff-recursion}) and (\ref{cyclic-recursion}) are
solved, i.e. $u_{\ff,m}$, $w_{\ff,m}$ are parameterized (at least
implicitly) in the terms of $u_m,w_m$, $m=1,...,M$, and some extra
parameters, possible degrees of freedom of (\ref{ff-recursion})
and (\ref{cyclic-recursion}) (e.g. $\lambda_\ff$, $\kappa_\ff$,
etc.) Then (\ref{uw-prime}) defines in general the transformation
$\mathcal{Q}_\ff$
\begin{equation}\label{Q-mapping}
\ds \mathcal{Q}_\ff\;:\;\{u_m^{},w_m^{}\}_{m=1}^M\;
\mapsto\;\{u_m',w_m'\}_{m=1}^M\;.
\end{equation}
The transfer matrices for two sets $\{u_m^{},w_m^{}\}$ and
$\{u_m',w_m'\}$ have the same spectrum, because of the existence of
$N^M\times N^M$ matrix $\Q_\ff$, nondegenerative in general,
\begin{equation}\label{R-monodromy}
\ds\Q_\ff\;\stackrel{def}{=}\;
\textrm{tr}_{\ff}\;\widehat{\Q}_{\ff}^{}\;,\;\;\;
 \widehat{\Q}_\ff^{}\;\stackrel{def}{=}\;
\R_{1,\ff}^{}\,\R_{2,\ff}^{}\,\dots\,\R_{M,\ff}^{}\;,
\end{equation}
such that
\begin{equation}\label{t-Q-permutation}
\ds \tp(\lambda\,;\{u_m^{},w_m^{}\})\,\cdot\, \Q_\ff\;=\;
\Q_\ff\,\cdot\, \tp(\lambda\,;\{u_m',w_m'\})\;.
\end{equation}
Subscript $\ff$ of $\Q_\ff$ stands as the reminder for the
parameters, arising in the solution of recursion, including at
least the spectral parameter $\lambda_\ff$, so $\Q_\ff$ means in
particular $\Q(\lambda_\ff)$. It is important to point out ones more
that $\Q_\ff$ ``remembers'' about its initial
$\{u_m^{},w_m^{}\}_{m=1}^M$ and its final $\{u_m',w_m'\}_{m=1}^M$
sets of the parameters.
In terms of the eigenvectors of the transfer matrix $\tp(\lambda)$
(\ref{t-eigen}), operator $\Q_\ff$ can be written as
\begin{equation}\label{Q-decomp}
\ds \Q_\ff\;=\;\sum_{t}\; |\Psi_t^{}\rangle\,q_{t,\phi}\,
\langle\Psi_t'|\;.
\end{equation}

Consider now the repeated application of the transformations
$\mathcal{Q}_\ff$, eq. (\ref{Q-mapping}),
\begin{equation}\label{sequence}
\ds \{u_m^{},w_m^{}\}\stackrel{\mathcal{Q}_{\ff_1}}{\mapsto}\{u_
m',w_m'\} \mapsto ... \{u_m^{(n-1)},w_m^{(n-1}\}
\stackrel{\mathcal{Q}_{\ff_n}}{\mapsto} \{u_m^{(n)},w_m^{(n)}\}
\mapsto ... \{u_m^{(g)},w_m^{(g)}\}\;,
\end{equation}
such that the set of isospectral quantum transfer matrices
\begin{equation}\label{transfer-set}
\ds \tp^{(n)}(\lambda) \;=\;
\tp(\lambda,\kappa;\{u_m^{(n)},w_m^{(n)}\}_{m=1}^M)\;,\;\;\;
n=0,...,g
\end{equation}
has arisen. Sequence (\ref{sequence}) defines the transformation
$\mathcal{K}$,
\begin{equation}\label{K-mapping}
\mathcal{K}^{(g)}\;:\; \{u_m^{}\equiv u_m^{(0)},w_m^{}\equiv
w_m^{(0)}\}_{m=1}^M\; \mapsto\;\{u_m^{(g)},w_m^{(g)}\}_{m=1}^M\;,
\end{equation}
with the finite dimensional counterpart
\begin{equation}\label{K-operator}
\ds\mathbf{K}^{(g)} \;=\;
\Q^{(1)}_{\phi_1}\,\Q^{(2)}_{\phi_2}\,...\,\Q^{(n)}_{\phi_n}\,...\,\Q^{(g)}_{\phi_g}
\;,
\end{equation}
where
\begin{equation}\label{Q-local}
\ds \Q^{(n)}_{\phi_n}\;=\; \textrm{tr}_{\phi_n}\;
\left(\R^{(n)}_{1,\ff_n}\,\R^{(n)}_{2,\ff_n}\,\dots\,
\R^{(n)}_{M,\ff_n}\right)\;,
\end{equation}
such that
\begin{equation}
\ds \tp^{(n-1)}(\lambda)\;\Q^{(n)}_{\phi_n}\;=\;
\Q^{(n)}_{\phi_n}\;\tp^{(n)}(\lambda)\;\;\; \textrm{and}\;\;\; \ds
\tp^{(0)}(\lambda)\;\mathbf{K}^{(g)}\;=\;
\mathbf{K}^{(g)}\;\tp^{(g)}(\lambda)\;.
\end{equation}
Thus in general the transformations $\mathcal{Q}_\ff$ and their
iterations $\mathcal{K}^{(g)}$ give the isospectral
transformations of the initial quantum chain in the space of the
parameters, so that their finite dimensional counterparts $\Q_\ff$
and $\mathbf{K}^{(g)}$ respectfully describe the change of the
eigenvector basis (see eq. (\ref{K-decomposition}).

\section{Matrix $\R$}

In this section we construct explicitly the finite dimensional
matrix $\R_{m,\ff}$, obeying (\ref{op-recursion}). Since a basis invariant
formula for $\R_{m,\ff}$ is useless and rather complicated,
we will find matrix elements of $\R_{m,\ff}$ in the basis
(\ref{representation}). But first we have to introduce several
notations concerning the functions on the Fermat curve.

\subsection{$w$-function}

Let $p$ be a point on the Fermat curve $\mathcal{F}$
\begin{equation}\label{fermat}
\ds
p\;\stackrel{def}{=}\;(x,y)\;\in\mathcal{F}
\;\;\Leftrightarrow\;\;
x^N+y^N=1\;.
\end{equation}
Actually, the identity (\ref{fc-gen}) is the origin of the Fermat
curve. Very useful function on the Fermat curve is $w_p(n)$,
$p\in\mathcal{F}$, $n\in\mathbb{Z}_N$, defined as follows:
\begin{equation}\label{woshka}
\ds {w_p(n)\over w_p(n-1)}\;=\;{y\over
1\,-\,x\,\omega^n}\;,\;\;\;w_p(0)\;=\;1\;.
\end{equation}

Function $w_p(n)$ has a lot of remarkable properties, see the
Appendix of ref. \cite{mss-vertex} for an introduction into
$\omega$-hypergeometry. In this paper it is necessary to mention
just a couple of properties of $w$-function. Let $O$ be the
following automorphism of the Fermat curve:
\begin{equation}\label{O}
\ds p\;=\;(x,y) \;\; \Leftrightarrow \;\;
Op\;=\;(\omega^{-1}x^{-1},\omega^{-1/2}x^{-1}y)\;.
\end{equation}
Then
\begin{equation}\label{Phi}
\ds w_{p}(n)\,w_{Op}(-n)\,\Phi(n) \;=\;1\;,\;\;\;
\textrm{where}\;\;\; \Phi(n)\;=\;(-)^n\,\omega^{n^2/2}\;.
\end{equation}
In the subsequent sections we will use also two simple
properties of $w$-function:
\begin{equation}\label{w-simp}
\ds w_{(x,\omega y)}(n)=\omega^n w_{(x,y)}(n)\;,\;\;\;
w_{(x,y)}(n+1)={y\over 1-\omega x} w_{(\omega x,y)}(n)\;.
\end{equation}
Define also three special points on the Fermat curve:
\begin{equation}\label{singular-p}
\ds q_0=(0,1)\;,\;\; q_\infty = Oq_0\;,\;\; q_1 =
(\omega^{-1},0)\;.
\end{equation}
Then
\begin{equation}\label{singular-w}
\ds w_{q_0}(n)=1\;,\;\; w_{q_\infty}(n) = {1\over\Phi(n)}
\;,\;\;
{1\over w_{q_1}(n)}\;=\;\delta_{n,0}\;.
\end{equation}

The inversion relation may be mentioned for the completeness
\begin{equation}\label{inversion}
\ds\sum_{n\in\mathbb{Z}_M}\; {w_{(\omega x,\omega y)}(n+a)\over
w_{(x,y)}(n+b)}\;=\;N\,{1-\omega x\over \omega x}\, {x^N\over
1-x^N}\, \delta_{a,b}\;.
\end{equation}

\subsection{Matrix elements of $\R_{m,\ff}$}

Consider the $N^2\times N^2$ matrix $\R_{m,\ff}(p_1,p_2,p_3)$ with
the following matrix elements:
\begin{equation}\label{R}
\ds\begin{array}{l}\ds
\langle\alpha_m,\alpha_\ff|\R_{m,\ff}|\beta_m,\beta_\ff\rangle\;=\\
\\
\ds =\; \omega^{(\alpha_m-\beta_m)\beta_\ff}\,
{w_{p_1}(\alpha_\ff-\alpha_m)\,w_{p_2}(\beta_\ff-\beta_m)\over
w_{p_3}(\beta_\ff-\alpha_m)}\;\delta_{\alpha_\ff,\beta_m}\;.
\end{array}
\end{equation}
Here $p_1,p_2,p_3$ are three points on the Fermat curve,
such that
\begin{equation}\label{xxxx}
\ds x_1\,x_2\;=\;x_3\;.
\end{equation}
Eq. (\ref{xxxx}) and the spin structure of (\ref{R}) provide the
dependence of (\ref{R}) on two continuous parameters, say $x_1$
and $x_3$, and on two discrete parameters, say the phase of $y_1$
and the phase of $y_3$.

\begin{prop}\label{prop-R}
Matrix $\R_{m,\ff}(p_1,p_2,p_3)$, whose matrix elements (\ref{R})
are given in the basis (\ref{representation}), makes the
following mapping:
\begin{equation}\label{finite-mapping}
\ds\left\{
\begin{array}{l}
\ds\R_{m,\ff}^{}\xop_m^{}\R_{m,\ff}^{-1}\;=\; \xop_\ff\;,\\
\\
\ds\R_{m,\ff}^{}\zop_m^{}\R_{m,\ff}^{-1}\;=\; {y_3\over
y_2}\zop_m^{}\zop_\ff^{}-\omega
{x_3y_1\over x_1y_2}\,\xop_\ff^{-1}\zop_\ff^{}\;,\\
\\
\ds\R_{m,\ff}^{}\xop_\ff^{-1}\R_{m,\ff}^{-1}\;=\; \omega
x_3\xop_m^{-1}-\omega{x_1y_3\over
y_1}\xop_m^{-1}\zop_m^{}\xop_\ff^{} +{y_3\over
y_1}\zop_m^{}\;,\\
\\
\ds \R_{m,\ff}^{}\zop_\ff^{-1}\R_{m,\ff}^{-1}\;=\; {y_3\over
y_2}\xop_\ff^{}\xop_m^{-1}- \omega {x_3y_1\over
x_1y_2}\zop_m^{-1}\xop_m^{-1}\;.
\end{array}\right.
\end{equation}
\end{prop}
\noindent\emph{Proof}. Each relation of (\ref{finite-mapping}) should be rewritten
in the form
\begin{equation}
\ds \langle\alpha_m,\alpha_\ff|\R_{m,\ff}\,\mathbf{f} \; = \;
\mathbf{f}'\, \R_{m,\ff}|\beta_m,\beta_\ff\rangle\;.
\end{equation}
Such identities may be verified directly with the help of
(\ref{w-simp}).\hfill $\Box$

Compare (\ref{finite-mapping}) with (\ref{the-mapping}).
$\R_{m,\ff}$ solves (\ref{op-recursion}) if
\begin{equation}\label{p-uw-1}
\ds x_{1}\;=\;\omegg\,{u_{\ff,m}\over\kappa\,u_m}
\;,\;\;\;{x_{3}\,y_{1}\over x_{1}\,y_{3}}\;=\;
\omegg\,{\lambda_\ff^{}\over u_{\ff,m} w_m}\;,
\end{equation}
and
\begin{equation}\label{p-uw-2}
\ds\begin{array}{ll} \ds
u_m'\;=\;{\kappa_\ff\over\kappa}\,u_{\ff,m}^{}\;,& \ds
w_m'\;=\;w_m^{}w_{\ff,m}^{}\,{y_{2}\over y_{3}}\;,\\
\\
\ds u_{\ff,m+1}^{}\;=\;\omega\,x_{3}\,u_m^{}\;, & \ds
w_{\ff,m+1}^{}\;=\;{\kappa\over\kappa_\ff}\,{u_m\over
u_{\ff,m}}\,{y_{3}\over y_{2}}\;.
\end{array}
\end{equation}
(\ref{p-uw-1}) and (\ref{p-uw-2}) are the complete set of the
relations following from the identification of
(\ref{finite-mapping}) with (\ref{the-mapping}). Using
(\ref{p-uw-1}), (\ref{xxxx}) and (\ref{fermat}), one has to fix
first the parameters $p_1$, $p_2$, $p_3$ and thus to define
$\R_{m,\ff}$. Then (\ref{p-uw-2}) parameterize $u_m'$, $w_m'$,
$u_{\ff,m+1}$, $w_{\ff,m+1}$ in the terms of $u_m$, $w_m$,
$u_{\ff,m}$, $w_{\ff,m}$ and the phases of $x_3$ and $y_2/y_3$.

When one considers the complicated operator $\mathbf{K}^{(g)}$,
eq. (\ref{K-operator}), its matrix elements must be calculated
using matrices $\R^{(n)}_{m,\ff_n}$, entering
(\ref{Q-local}). The matrix elements of all
$\R^{(n)}_{m,\ff_n}$ are the same functions,
\begin{equation}\label{R-nm}
\ds \R^{(n)}_{m,\ff_n} \;=\;
\R^{}_{m,\ff_n}(p_{1,m}^{(n)},p_{2,m}^{(n)},p_{3,m}^{(n)})\;,
\end{equation}
given by (\ref{R}), but with different and rather complicated
parameters. Eqs. (\ref{p-uw-1}) and (\ref{p-uw-2}) are written for
$p_{1,m}^{(1)}$, $p_{2,m}^{(1)}$, $p_{3,m}^{(1)}$ in fact.
Nevertheless, for given $\mathcal{K}^{(g)}$, i.e. for given sets
of $\{u_m^{(n)},w_m^{(n)}\}$, $m=1,...,M$ and $n=0,...,g$, see
(\ref{sequence}), one may parameterize the corresponding
$\mathbf{K}^{(g)}$ explicitly via parameterizations
(\ref{p-uw-1},\ref{p-uw-2}), the form  of the matrix elements of
$\R_{m,\ff}$ (\ref{R}) and formulas
(\ref{Q-local},\ref{K-operator}).

\subsection{Normalization of the matrix $\R$}

The matrix elements of $\R_{m,\ff}$ are defined in general up to a constant
multiplier. Definition (\ref{woshka}) and formula
(\ref{R}) are the most simple expressions in the terms of the
Fermat curve coordinates $p_1,p_2,p_3$. But this $\R_{m,\ff}$ is
not normalized. For several applications we need the
normalization of $\R_{m,\ff}$ connected with the
determinant of $\R_{m,\ff}$.

To calculate the determinant, we need several definitions. Let
\begin{equation}\label{V}
\ds V(x)\;\stackrel{def}{=}\;\prod_{n=1}^{N-1}\,
(1\,-\,\omega^{n+1} x)^n\;.
\end{equation}
$V(x)$ obeys the following relations:
\begin{equation}
\ds {V(\omega^{-1}x)\over V(x)}\;=\; {1-x^N\over (1-x)^N}\;,
\end{equation}
and
\begin{equation}
\ds V(\omega^{-1}x^{-1})\;=\; {\ds
(1-x^N)^{N-1}\omega^{N(N-1)(2N-1)/6}\over
(-x)^{N(N-1)/2}\,V(x)}\;.
\end{equation}
Besides, one may calculate the particular value:
\begin{equation}
V(\omega^{-1})\;\equiv\;N^{N/2}\,\EXP^{i\pi(N-1)(N-2)/12}\;.
\end{equation}
Function $V(x)$ appears in the following expressions, where
$p=(x,y)$:
\begin{equation}\label{w-products}
\ds \begin{array}{l} \ds
\prod_n\sum_\sigma\,{\omega^{n\sigma}\over w_p(\sigma)} \;=\;
(\omega x)^{N(N-1)/2}\,{V(\omega^{-1})\over
V(x)}\;,\\
\\
\ds \prod_n\sum_\sigma\,{\omega^{n\sigma}\over \Phi(\sigma)
w_p(\sigma)} \;=\; {V(\omega^{-1})\over
V(x)}\;,\\
\\
\ds \prod_n w_p(n)\;=\;{V(x)\over y^{N(N-1)/2}}\;.
\end{array}
\end{equation}
The third expression is rather trivial. One may prove the first
two formulas considering the poles and zeros of the left and right
hand sides. Details may be found in the appendix of
\cite{mss-vertex}.

\begin{prop}\label{prop-R-det}
The determinant of the matrix $\R$, defined by (\ref{R}), is
\begin{equation}\label{R-det}
\ds  \det \R\;=\; (-)^{N(N-1)/2}\, \left(\prod_n{w_{p_2}(n)\over
w_{Op_1}(n)}\right)^N\,\left(\prod_n\sum_\sigma\,
{\omega^{n\sigma}\over\Phi(\sigma) w_{p_3}(\sigma)}\right)^N\;.
\end{equation}
\end{prop}
\noindent\emph{Proof}. To obtain (\ref{R-det}), one has to use $O$-automorphism
(\ref{Phi})  for $w_{p_1}$ in (\ref{R}). The factors $w_{p_2}$ and
$w_{Op_1}$ correspond to the diagonal matrices $D$ and $D'$ in the
matrix decomposition of $\R_{m,\ff}$,
\begin{equation}
\ds \R_{m,\ff}\;=\;D(p_1) R'(p_3) D'(p_2)\;.
\end{equation}
The determinants of $D$ and $D'$ give the term with $p_1$ and
$p_2$ in (\ref{R-det}). The Fourieur transform
applied to the matrix $\R_{m,\ff}$
in the $m$-th space yields
$F_m^{} R'(p_3) F_m^{-1} = D''(p_3) P_{m,\ff}$, where
$\langle\alpha_m|F|\beta_m\rangle=N^{-1/2}\omega^{\alpha_m\beta_m}$,
$D''$
is a diagonal matrix and $P_{m,\ff}$ is the permutation.
$\det{D''}$ gives the term, depending on $p_3$ in (\ref{R-det}),
and $\det P$ is the sign factor in (\ref{R-det}). \hfill $\Box$

Let the normalization factor for $\R_{m,\ff}$ be $\rho_R$:
\begin{equation}
\ds\rho_R^N\;=\;\left(\omega^{-1/2}{y_1\over
x_1y_2}\right)^{N(N-1)/2}\, {V(x_1)\over V(\omega^{-1}x_2^{-1})
V(x_3)}\;,
\end{equation}
so that
\begin{equation}
\ds\det\R\;=\;\textrm{const}\,\cdot\,\rho_R^{N^2}\;.
\end{equation}
Analogously, the normalization factor for the monodromy of
$\R_{m,\ff}$ and so of $\Q_\ff$ is
\begin{equation}\label{Q-norm}
\ds\rho_Q^N\;=\;\prod_{m=1}^M\;\left(\omega^{-1/2}{y_{1,m}\over
x_{1,m}y_{2,m}}\right)^{N(N-1)/2}\, {V(x_{1,m})\over
V(\omega^{-1}x_{2,m}^{-1}) V(x_{3,m})}\;.
\end{equation}

\section{Parameterization of the recursion}

The main object of the present investigations is the system of the
recursion relations (\ref{ff-recursion}). Our goal is the
construction of operator $\mathbf{K}^{(g)}$, (\ref{K-operator}),
\emph{for the homogeneous initial state} (\ref{uw-homo-3}). In
this case, the fist step transformation $\mathcal{Q}_{\ff_1}$ has
a remarkably simple structure, clarifying nevertheless the
structure of all subsequent $\mathcal{Q}_{\ff_n}$.

\subsection{The first step}

Let the initial parameters $u_m,w_m$ are homogeneous
\begin{equation}\label{uw-homo-3}
\ds u_m\;=\;-\omega^{-1/2}\;,\;\;w_m=-1\;,
\end{equation}
see (\ref{uw-homo}). The main recursion relation is the first
equation in (\ref{ff-recursion}). For the homogeneous initial
state it contains only one unknown $u_{\ff,m}$. Without loss of
generality let us introduce complex numbers $\delta_\ff^{}$ and
$\delta_\ff^*$ ($*$ does not stand for the complex conjugation!)
and a function $\tau'_m$, $m\in\mathbb{Z}$, such that
\begin{equation}\label{uphim}
\ds
u_{\ff,m}\;=\;-\omega^{-1/2}\,\delta_\ff^{}\,{\tau'_{m-1}\over\tau'_m}\;,
\;\;\;\delta_\ff^*\;=\;\frac{\lambda_\ff^{}}{\delta_\ff}\;.
\end{equation}
The first relation in (\ref{ff-recursion}) may be rewritten
as the second order linear recursion for
$(\tau'_m)^N$:
\begin{equation}\label{linear-recursion}
\ds\delta_\ff^N\delta_\ff^{*N}\,\left(\tau'_{m}\right)^N\,=\,
\delta_\ff^N\,\left(\tau'_{m-1}\right)^N \,+\,
\delta_\ff^{*N}\,\left(\tau'_{m+1}\right)^N \,+\,
\kappa^N\,\left(\tau'_m\right)^N\;.
\end{equation}
Parameterization (\ref{uphim}) and linear recursion
(\ref{linear-recursion}) are considered now in the general
position, without mentioning $\mathbb{Z}_M$-invariance. It is
known that the general solution of a second order linear recursion
with constant coefficients is a linear combination of two
fundamental solutions, $(\tau_m')^N=\alpha z_1^m+\beta z_2^m$,
where $z_1$ and $z_2$ are two roots of the characteristics
equation $(\delta_\ff^{}\delta_\ff^*)^N=\delta_\ff^N
z^{-1}+\delta_\ff^{*N} z + \kappa^N$. One may always redefine
$(\tau_m')^N\mapsto\alpha^{-1}z_1^{-m}(\tau_m')^N$. This
corresponds to a redefinition of $\delta_\ff$. One may choose
$\delta_\ff$
\begin{equation}\label{delta-NN}
\ds \delta_\ff^N\delta_\ff^{*N}\,=\,
\delta_\ff^N\,+\,\delta_\ff^{*N}\,+\,\kappa^N\;,\;\;\;
\lambda_\ff^{}\;=\;\delta_\ff^{}\delta_\ff^{*}\;,
\end{equation}
such that $z_1=1$ and $\tau'_m=1$ solves (\ref{linear-recursion}).
Compare (\ref{delta-NN}) with equations
(\ref{phi-parametrization}), (\ref{Delta-curve}).
One may identify
\begin{equation}\label{delta-Delta}
\ds
\delta_\ff^N\;=\;\Om_\ff^{}\,\;\;\;\delta_\ff^{*N}\;=\;\Om_\ff^{*N}\;,\;\;\;
\lambda_\ff^N\;=\;\Lambda_\ff^{}\;,
\end{equation}
so the index $\phi$ may be understood conveniently as the spectral
parameter $\phi$.

The second fundamental solution of the characteristics equation is
$z_2=(\delta_\ff/\delta_\ff^*)^N$, i.e.
$(\tau'_m)^N=(\delta_\ff^{N}/\delta_\ff^{*N})^m$. The
complete solution of recursion (\ref{linear-recursion}) is
\begin{equation}\label{tau-prime}
\ds \left(\tau'_m\right)^N\;=\;1\;-\;f_\ff\,\EXP^{2 i m \phi}\;,
\end{equation}
where $f_\ff$ is arbitrary complex number, and (see
(\ref{phi-parametrization}))
\begin{equation}
\ds
\EXP^{2\,i\,\phi}\;=\;{\delta_\ff^N\over\delta_\ff^{*N}}\;\equiv\;{\Om_\ff^{}\over\Om_\ff^*}\;.
\end{equation}

Turn now to the $\mathbb{Z}_M$-invariance condition. It demands
$\tau'_{M+m}=\tau_m'$, hence the possible values of $\phi$ become
discrete, $\ds\EXP^{2i\phi M}=1$ and $\EXP^{2i\phi}\neq
1$. Since there exists the symmetry
between $\delta_\ff^{}$ and $\delta_\ff^*$ in (\ref{linear-recursion}),
we may regard $\phi\in\Set$ in (\ref{tau-prime}), see (\ref{phi-k}) for the
definition of $\Set$.

Turn further to the first relation in (\ref{uw-prime}) (or to the
first relation in (\ref{p-uw-2})). It follows
\begin{equation}
\ds
u_m'\;=\;-\omega^{-1/2}\,{\kappa_\ff\over\kappa}\,\delta_\ff^{}\,{\tau'_{m-1}\over\tau'_m}\;.
\end{equation}
Taking into account the gauge invariance (\ref{gauge-invariance}),
we may fix $\kappa_\ff$ as follows
\begin{equation}\label{kappa-phi}
\ds\kappa_\ff\;=\;{\kappa\over\delta_\ff}
\end{equation}
and so
\begin{equation}\label{u-prime}
\ds u_m'\;=\;-\omega^{-1/2}\,{\tau'_{m-1}\over\tau'_m}\;.
\end{equation}
It agrees with (\ref{gauge-fix}). The second equations in
(\ref{ff-recursion}) and (\ref{uw-prime}) give
\begin{equation}\label{w-prime}
\ds w_m'\;=\;-\,{\theta'_m\over\theta'_{m-1}}\;,
\end{equation}
where
\begin{equation}\label{theta-prime}
\ds \left(\theta_m'\right)^N\;=\;1\,-\,f_\ff\,
{\Om^*_{\phi}\over\Om^{}_{\phi}}\; {\Om^{}_{\phi}\,-\,1\over
\Om^*_{\phi}\,-\,1}\,\EXP^{2i m \phi}\;.
\end{equation}

These simple calculations show that with the useful
parameterizations (\ref{delta-Delta}), where the functions of
$\phi$ are defined in (\ref{phi-parametrization}), and with
additional condition (\ref{kappa-phi}), the image of
$\mathcal{Q}_\ff$-transformation
(\ref{tau-prime},\ref{u-prime},\ref{w-prime},\ref{theta-prime}),
contains one arbitrary complex parameter $f_\ff$ if $\phi\in\Set$.
The case when $-\phi\in\Set$ is equivalent to the case when
$\phi\in\Set$. It follows from the symmetry between
$\delta_\ff^{}$ and $\delta_\ff^*$. The case when $\phi$ is
arbitrary complex number, $\mathbb{Z}_M$-invariance requirement
yields $f_\ff\equiv 0$, therefore $\tau_m'\;=\;1$ and the chain
remains the homogeneous one.

Equations (\ref{tau-prime}) and (\ref{theta-prime}) define the
$N$th powers of $\tau'$ and $\theta'$. Their phases are arbitrary.
The $N$th rooting is the subject of the definition of
$p_{j,m}$ for $\R_{m,\ff}$.

Functions $(\tau'_m)^N$ and $(\theta'_m)^N$ look like solutions of
a Hirota-type discrete equation. We call these expressions
``one-soliton $\tau$-functions''. Therefore $\mathcal{Q}_\ff$-transformation with
$\phi\in\Set$ must be identified with the B\"acklund transformation for
the classical relativistic Toda chain, see \cite{Sklyanin-rev} for
the application of this ideology to the usual Toda chain.

\subsection{$N$th step}

Turn now to the sequence (\ref{sequence}) of
$\mathcal{Q}_{\ff_n}$-transformations
\begin{equation}
\ds  \mathcal{Q}_{\ff_n}\;:
\;\{u_m^{(n-1)},w_m^{(n-1)}\}_{m=1}^M\;\mapsto\;\{u_m^{(n)},w_m^{(n)}\}_{m=1}^M
\end{equation}
with generic $\lambda_{\ff_n}$, parameterized according to
(\ref{delta-Delta},\ref{phi-parametrization}) as follows
\begin{equation}\label{delta-Delta-ones-more} \ds
\begin{array}{l} \ds \delta_n^N\,=\,\Om_n^{}\;,\;\;\;
\delta_n^{*N}\,=\,\Om_n^*\;,\;\;\;\lambda_{\ff_n}^{}\,=\,\delta_n^{}\delta_n^*\;,\\
\\
\ds \Om_n^{}\Om_n^*\,=\,\Om_n^{} \,+\, \Om_n^* \,+\,
\kappa^N\;,\;\;\; \EXP^{2 i \phi_n} \,=\,
{\Om_n^{}\over\Om_n^*}\;.
\end{array}
\end{equation}
Taking into account (\ref{gauge-fix}), we may parameterize the
$n$-th state in the sequence (\ref{sequence}), $n=0,...,g$, as
\begin{equation}\label{uw-inhomo}
\ds u_m^{(n)} \;=\; -\omega^{-1/2}\,{\tau^{(n)}_{m-1}\over\tau^{(n
)}_m}\;, \;\;\;
w_m^{(n)}\;=\;-\,{\theta^{(n)}_m\over\theta^{(n)}_{m-1}}\;, \;\;\;
m\,\in\,\mathbb{Z}_M\;.
\end{equation}
The homogeneous initial state (\ref{uw-homo}) corresponds
to $\tau^{(0)}_m = \theta^{(0)}_m = 1$. Parameterizations
(\ref{uw-inhomo}) with the first relation of (\ref{uw-prime}) taken into
account implies that
\begin{equation}\label{kappa-phi-n}
\ds \kappa_{\ff_n}\;=\;{\kappa\over\delta_{n}}
\end{equation}
and
\begin{equation}\label{u-phi-m}
\ds u_{\ff_n,m} \;=\; -
\omega^{-1/2}\,\delta_{n}\,{\tau^{(n)}_{m-1}\over\tau^{(n)}_{m}}
\end{equation}
are imposed for $\mathcal{Q}_{\ff_n}$-transformation. The key
recursion relation, the first one from (\ref{ff-recursion}), looks
now as follows
\begin{equation}\label{triplet}
\begin{array}{l}
\ds \left(\delta_{\ff_n}^{}\delta_{\ff_n}^{*}\, \tau^{(n)}_{m}\,
\tau^{(n-1)}_m\, \theta^{(n-1)}_{m-1}\right)^N \,=\,
\left(\delta_{\ff_n}^{}\, \tau^{(n)}_{m-1}\, \tau^{(n-1)}_{m}\,
\theta^{(n-1)}_{m}\right)^N \\
\\
\ds +\, \left(\delta_{\ff_n}^{*}\, \tau^{(n)}_{m+1}\,
\tau^{(n-1)}_{m-1}\, \theta^{(n-1)}_{m-1}\right)^N \,+\,
\left(\kappa\, \tau^{(n)}_m\, \tau^{(n-1)}_{m-1}\,
\theta^{(n-1)}_m\right)^N\;.
\end{array}
\end{equation}
The second equations in (\ref{ff-recursion}) and (\ref{uw-prime})
provide
\begin{equation}\label{w-phi-m}
\ds w_{\ff_n,m+1}^{}\;=\;w_{\ff_n}\,
{\theta^{(n-1)}_m\,\tau^{(n)}_m
\over\tau^{(n-1)}_m\,\theta^{(n)}_m}\;, \;\;\;
w_{\ff_n}^N\;=\;{1\over 1\,-\,\Om^*_{\ff_n}}\;,
\end{equation}
and
\begin{equation}\label{theta-rec}
\ds \left(\theta^{(n)}_m\right)^N\;=\;
\left({w_{\ff_n}\over\tau^{(n-1)}_{m-1}}\right)^N\,
\left(\left(\theta^{(n-1)}_m\tau^{(n)}_{m-1}\right)^N\,-\,
\left(\delta_{\ff_n}^{*}\theta^{(n-1)}_{m-1}
\tau^{(n)}_{m}\right)^N\right)\;.
\end{equation}

Consider (\ref{triplet}) and (\ref{theta-rec}) as the set of discrete
equations with respect to $m\in\mathbb{Z}$ and $n\in\mathbb{Z}_+$ with
the initial data $\tau_m^{(0)}=\theta_m^{(0)}=1$.
\begin{prop}\label{prop-inhomo}
General solution of the infinite dimensional in $m$ and in $n$
system of discrete equations (\ref{triplet}), (\ref{theta-rec})
with generic values
of $\phi_n$, $n=0,1,...$ (see parameterizations
(\ref{delta-Delta-ones-more}) and with the initial data
$\tau_m^{(0)}=\theta_m^{(0)}=1$ is given by
\begin{equation}\label{tau-and-theta}
\ds \begin{array}{l} \ds \left(\tau_m^{(n)}\right)^N\;=\;
h^{(n)}(\{f_k\EXP^{2 i m \phi_k}\}_{k=1}^n)\;,\\
\\
\ds \left(\theta_m^{(n)}\right)^N\;=\; h^{(n)}(\{f_ks_k(1)\EXP^{2
i m \phi_k}\}_{k=1}^n)\;,
\end{array}
\end{equation}
where $f_1,...,f_n\equiv\{f_k\}_{k=1}^n$ is a set of arbitrary
complex variables, $h^{(n)}$ are defined by
\begin{equation}
\ds h^{(1)}(f_1)\;=\; 1\,-\,f_1\;,
\end{equation}
and recursively
\begin{equation}\label{h-definition}
\ds\begin{array}{lll} \ds  h^{(n)}(\{f_k\}_{k=1}^n) & = & \ds
h^{(n-1)}(\{f_k s_k(\Om_n^{*})\}_{k=1}^{n-1})
\\
\\
\ds & - & \ds f_n\,\prod_{m=1}^{n-1}\, s_n(\Om_m^{*})\;
h^{(n-1)}(\{f_k s_k(\Om_n^{})\}_{k=1}^{n-1})\;,
\end{array}
\end{equation}
where functions $s_n$ are
\begin{equation}\label{s-function}
\ds s_n(\xi)\;=\; {\Om_n^{*}\over \Om_n^{}}\; {\Om_n^{}\,-\,\xi
\over \Om_n^{*} \,-\,\xi}\;.
\end{equation}
\end{prop}
\noindent\emph{Proof}. Equations (\ref{triplet}) and
(\ref{theta-rec}) become the algebraic identities after the
substitution (\ref{tau-and-theta}) (see Proposition
\ref{prop-baecklund} on page \pageref{prop-baecklund} of the
Appendix). The first step of this recursion with respect to $n$ is
described before, hence $(\tau_m')^N=h^{(1)}(f\EXP^{2i\phi m})$
etc.

Suppose that the statement of the Proposition is true for
$n=1,...,g-1$. $h^{(g)}$ depends on $f_1,...,f_g$, while
$h^{(g-1)}$ depends only on $f_1,...,f_{g-1}$, therefore
$\tau_m^{(g)}$ contains one extra degree of freedom
$f_g\in\mathbb{C}$  with respect to $\tau_m^{(g-1)}$,
$\theta_m^{(g-1)}$. On the other side, (\ref{triplet}) with $n=g$
is the second order linear recursion with respect to
$(\tau_m^{(g)})^N$, so it must have exactly two fundamental
solutions. Hence, due to the homogeneity, general normalized
solution of (\ref{triplet}) with $n=g$ with respect to
$(\tau_m^{(g)})^N$ for given $\tau_m^{(g-1)}$, $\theta_m^{(g-1)}$
must have exactly one extra degree of freedom. This degree of the
freedom is parameter $f_g$.\hfill $\Box$

Proposition (\ref{prop-inhomo}) is formulated for the generic
sequence of $\phi_n$. One may verify, due to definition
(\ref{h-definition}), $h^{(g)}(\{f_n\}_{n=1}^g)$ is the
symmetrical function with respect to any permutation of the pairs
$(f_n,\phi_n)$, $n=1,...,g$.

Two special cases should be discussed. The fist one is the case
when $\phi_n=\phi_{n'}\mod\pi$ for some $n$ and $n'$, and the
second one is $\phi_n=-\phi_n'\mod\pi$. Both these cases are the
singular ones for definition (\ref{h-definition}). Nevertheless,
since the ``observables'' $u_m^{(g)}$ and $w_m^{(g)}$ are the
ratios of $\tau$-functions, corresponding $\tau_m^{(g)}$ and
$\theta_m^{(g)}$ for these cases may be obtained as the residues
of formula (\ref{tau-and-theta}). Technically, both
cases are equivalent (up to some re-parameterizations of
$\kappa_{\phi_n}$ or $\kappa_{\phi_{n'}}$). The equivalence is
connected again with the symmetry between $\delta_\ff^{}$ and
$\delta_\ff^*$, so we may consider only the case
\begin{equation}\label{annihilation-phi}
\ds \phi_n\;=\;-\,\phi_{n'}\,\mod\pi\;.
\end{equation}
In this case condition (\ref{kappa-phi-n}) is convenient for any
$n$, and the residues of $\tau_m^{(g)}$ and $\theta_m^{(g)}$ are
given  by the same formula (\ref{tau-and-theta}), but with
\begin{equation}\label{annihilation-f}
\ds f_n\;=\;f_{n'}\;\equiv\;0\;.
\end{equation}

Turn now the the finite chain with the periodic boundary
conditions
\begin{equation}\label{tt-bc}
\ds \tau_{M+m}^{(g)}\;=\;\tau_{m}^{(g)}\;,\;\;\;
\theta_{M+m}^{(g)}\;=\;\theta_m^{(g)}\;.
\end{equation}
This boundary conditions, being applied to (\ref{tau-and-theta}),
give the following rule: in the sequence of
$f_1,...,f_g$
\begin{equation}
\ds f_n\;=\;0\;\;\;\Leftrightarrow\;\;\;
\phi_n\not\in\Set\mod\pi\;,
\end{equation}
where the set $\Set$ is given by (\ref{phi-k}). It means that any
$\mathcal{Q}_\ff$ with $\phi\not\in\Set$ just changes the values of
$f_n$. Therefore the maximal number of the possible exponents $\EXP^{2 i
m \phi_n}$ is $M-1$. Without loss of generality, let us define the
minimal complete sequence of $\mathcal{Q}_{\ff_n}$ described by
\begin{equation}\label{minimal-set}
\ds \{(f_n,\phi_n)\}_{n=1}^{M-1}\;\;:\;\;\phi_n\in\Set\;,\;\;\;
\phi_n\neq\phi_{n'}\;,
\end{equation}
such that the corresponding operator $\mathbf{K}^{(M-1)}$ creates the state
of the quantum chain with $(M-1)$ arbitrary complex parameters.

Usually the expressions for $(\tau_m^{(g)})^N$,
$(\theta_m^{(g)})^N$, given by (\ref{tau-and-theta}), are called
the solitonic solutions of discrete Hirota-type equations. $g$ is
the number of solitons, $\EXP^{2 i m \phi_n}$ is $n$th solitonic
wave (or exponent) and $f_n$ is the amplitude of $n$th partial
wave. $\mathcal{Q}_{\ff_n}$-mapping, $\phi_n\in\Set$, increases
the number of the solitons, hence it is the B\"acklund
transformation.

\subsection{Parameterization of modified $Q$-operators}

Turn now to the parameterizations of the matrix elements of the
finite dimensional counterpart of the transformation
$\mathcal{Q}_{\phi_n}$ -- the matrix $\Q_{\phi_n}$. By definition
(\ref{Q-local}), it is constructed with the help of
$\R$-matrices, and to parameterize it, one has to point out the
Fermat curve parameters (\ref{R-nm}).

All the ingredients are already prepared: one has just substitute
the parameterizations
(\ref{uw-inhomo},\ref{kappa-phi-n},\ref{u-phi-m},\ref{w-phi-m})
into (\ref{p-uw-1}) and (\ref{p-uw-2}). As the result,
$\R^{(n)}_{m,\ff_n}$-matrices, entering (\ref{Q-local}), have the
arguments (see (\ref{R-nm})):
\begin{equation}\label{xy-explicit}
\ds\begin{array}{ll} \ds x_{1,m}^{(n)}\;=\;
\omega^{-1/2}\,{\delta_n\over\kappa}\,
{\tau_{m\phantom{1}}^{(n-1)}\tau_{m-1}^{(n)}
\over\tau_{m-1}^{(n-1)}\tau_{m\phantom{1}}^{(n)}}\;,& \ds
x_{3,m}^{(n)}\;=\;\omega^{-1}\delta_n^{}\,
{\tau_{m\phantom{1}}^{(n-1)}\tau_{m\phantom{1}}^{(n)}\over
\tau_{m-1}^{(n-1)}\tau_{m+1}^{(n)}}\;,\\
&\\
\ds {y_{3,m}^{(n)}\over y_{2,m}^{(n)}}\;=\;w_{\phi_n}\,
{\theta_{m\phantom{1}}^{(n-1)}\tau_{m-1}^{(n)}\over
\tau_{m-1}^{(n-1)}\theta_{m\phantom{1}}^{(n)}}\;, & \ds
{y_{1,m}^{(n)}\over y_{3,m}^{(n)}}\;=\;
\omega^{1/2}\,{\lambda_n\over\kappa\delta_n}\,
{\theta_{m-1}^{(n-1)}\tau_{m+1}^{(n)}\over
\theta_{m\phantom{1}}^{(n-1)}\tau_{m\phantom{1}}^{(n)}}\;,
\end{array}\end{equation}
and due to (\ref{xxxx})
\begin{equation}
\ds x_{2,m}^{(n)}\;=\;
\omega^{-1/2}\kappa\,{(\tau_m^{(n)})^2\over\tau_{m-1}^{(n)}
\tau_{m+1}^{(n)}}\;.
\end{equation}

The trace of quantum monodromy (\ref{Q-local}) may be calculated
explicitly in the basis (\ref{representation})
\begin{equation}\label{Q-elements}
\ds \langle\alpha|\mathbf{Q}^{(n)}_{\phi_n}|\beta\rangle =
\prod_{m\in\mathbb{Z}_M}\;
\omega^{(\alpha_m-\beta_m)\beta_{m+1}}\,
{w_{p_{1,m}^{(n)}}(\beta_m-\alpha_m)\,
w_{p_{2,m}^{(n)}}(\beta_{m+1}-\beta_m)\over
w_{p_{3,m}^{(n)}}(\beta_{m+1}-\alpha_m)}\;.
\end{equation}
Thus, the operator $\mathbf{K}^{(g)}$, (\ref{K-operator}), calculated as the
product of $g$ $Q^{(n)}$-operators, is defined explicitly.

\subsection{Permutation of the modified $Q$-operators}

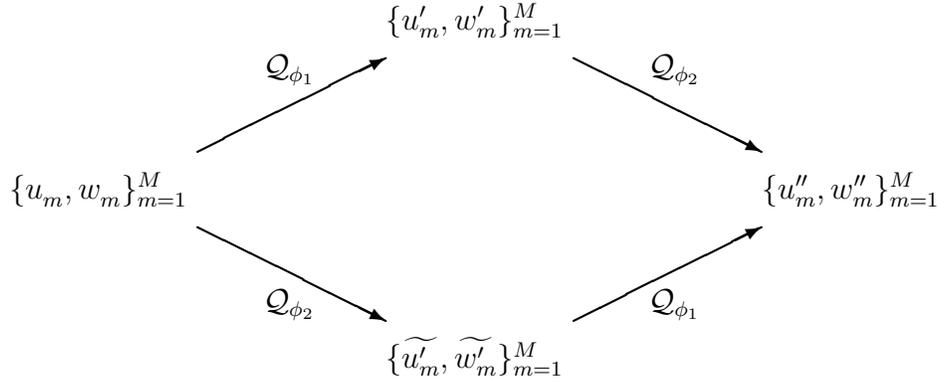
\begin{figure}
\begin{center}
\setlength{\unitlength}{0.25mm} % 150mm=600units is the width of the page.
\thicklines
\begin{picture}(500,200)
\put(0,90){$\{u_m^{},w_m^{}\}_{m=1}^M$}
\put(400,90){$\{u_m'',w_m''\}_{m=1}^M$}
\put(200,0){$\{\widetilde{u_m'},\widetilde{w_m'}\}_{m=1}^M$}
\put(200,180){$\{u_m',w_m'\}_{m=1}^M$}
\put(100,115){\vector(2,1){100}} \put(300,165){\vector(2,-1){100}}
\put(100,75){\vector(2,-1){100}} \put(300,25){\vector(2,1){100}}
\put(135,155){$\mathcal{Q}_{\phi_1}$}
\put(340,155){$\mathcal{Q}_{\phi_2}$}
\put(135,30){$\mathcal{Q}_{\phi_2}$}
\put(340,30){$\mathcal{Q}_{\phi_1}$}
\end{picture}
\end{center}
\caption{Two different ordering of $\mathcal{Q}_{\phi_1}$ and
$\mathcal{Q}_{\phi_2}$.} \label{fig-diagram}
\end{figure}

For a given state $\{u_m,w_m\}_{m=1}^M$ let us consider two successive
$\mathcal{Q}$-transformations (\ref{Q-mapping}) with different
$\phi$-parameters $\phi_1$ and $\phi_2$. For the simplicity let
$\{u_m,w_m\}_{m=1}^M$ be a solitonic state, described by
(\ref{uw-inhomo}) with (\ref{tau-and-theta}). These two
transformations may be applied in two different orders, but in our
definition of $h^{(g)}$ (\ref{h-definition}) as the
symmetrical functions of the pairs $(\phi_k,f_k)$ the result should
be the same. Of course, the intermediate states are different.
These two different ordering are shown in Fig.
(\ref{fig-diagram}).

In the terms of the transfer matrices and modified $\Q$-operators
the diagram means
\begin{equation}
\ds\left\{\begin{array}{l} \ds
\tp(\lambda;\{u_m,w_m\}_{m=1}^M)\;\Q_{\phi_1}^{}\;=\;
\Q_{\phi_1}^{}\;\tp(\lambda;\{u_m',w_m'\}_{m=1}^M)\;,\\
\\
\ds \tp(\lambda;\{u_m',w_m'\}_{m=1}^M)\;\Q_{\phi_2}^{}\;=\;
\Q_{\phi_2}^{}\;\tp(\lambda;\{u_m'',w_m''\}_{m=1}^M)\;,
\end{array}\right.
\end{equation}
versus
\begin{equation}
\ds\left\{\begin{array}{l} \ds \tp(\lambda;\{u_m,w_m\}_{m=1}^M)\;
\widetilde{\Q}_{\phi_2}^{}\;=\; \widetilde{\Q}_{\phi_2}^{}\;
\tp(\lambda;\{\widetilde{u_m'},\widetilde{w_m'}\}_{m=1}^M)
\;,\\
\\
\ds \tp(\lambda;\{\widetilde{u_m'},\widetilde{w_m'}\}_{m=1}^M) \;
\widetilde{\Q}_{\phi_1}^{}\;=\;
\widetilde{\Q}_{\phi_1}^{}\;\tp(\lambda;\{u_m'',w_m''\}_{m=1}^M)\;.
\end{array}\right.
\end{equation}
Since the result of both mappings is the same, two products of
\emph{different} $\Q$-operators should be the same up to a
multiplier. This multiplier corresponds to the normalization
factors (\ref{Q-norm})
\begin{equation}\label{Q-exchange}
\ds \rho_{Q_1}^{-1}\rho_{Q_2}^{-1}\;
\Q_{\phi_1}^{}\,\Q_{\phi_2}^{} \;=\; \rho_{\widetilde{Q}_2}^{-1}
\rho_{\widetilde{Q}_1}^{-1}\; \widetilde{\Q}_{\phi_2}^{}\,
\widetilde{\Q}_{\phi_1}^{}\;.
\end{equation}
This equation is provided actually by an intertwining
relation
\begin{equation}
\ds \rho_{R_1}^{-1}\rho_{R_2}^{-1}\;
\R_{m,\phi_1}^{}\,\R_{m,\phi_2}^{}\,\mathbf{S}_{\phi_1,\phi_2}^{}\;=\;
\rho_{R_1'}^{-1}\rho_{R_2'}^{-1}\;
\mathbf{S}_{\phi_1,\phi_2}'\,\R_{m,\phi_2}'\, \R_{m,\phi_1}'\;,
\end{equation}
where $\mathbf{S}$ is a sort of the Chiral Potts Model $R$-matrix.
These details are not essential in this paper and besides need a
lot of extra notations and definitions.

Eq. (\ref{Q-exchange}) provides the following
\begin{prop}\label{prop-ordering}
Operator $\mathbf{K}^{(M-1)}$, given by (\ref{K-operator}), and
parameterized with the help of $\tau_m^{(n)}$ and
$\theta_m^{(n)}$, defined for the minimal complete sequence
(\ref{minimal-set}), does not depend (up to a multiplier!) on the
ordering of the data $\{(f_n,\phi_n)\}_{n=1}^{M-1}$  Thus, without
loss of generality, one may fix the ordering in (\ref{K-operator})
and in $\Set$ as follows
\begin{equation}
\ds \phi_n\;=\;{\pi n\over M}\;,\;\;\; n=1,...,M-1\;.
\end{equation}
\end{prop}

\subsection{Inhomogeneous chain}\label{sec-inhomo}

In this section we investigated and solved the problem of the
isospectrality for the homogeneous initial state (\ref{uw-homo})
of the quantum chain. For the chain of the size $M$ with the
periodic boundary conditions the solution is described in the
terms of Hirota-type solitonic $\tau$-functions
(\ref{tau-and-theta}) (the number of independent
$f_n$ is equal to $M-1$).

Usual $\Q$-operators are the particular cases of our modified
$\Q$-operators. Namely, let us consider the first step of the recursion
in $n$ (\ref{linear-recursion}) ones more. If $\phi$ in
(\ref{tau-prime}) and (\ref{theta-prime}) is such that
$\phi\not\in\Set\mod\pi$ (this corresponds to a generic value of
$\lambda_\ff$) then $\mathbb{Z}_M$-periodicity demands $f\equiv
0$, and $\tau_m'=\theta_m'=1$. It means,
$\mathcal{Q}_\ff$-transformation is the trivial one, and the
corresponding $\Q_\ff$-operator changes nothing. Evidently, such
$Q$-operators $\Q(\lambda_\ff)$ commute with the transfer matrix $\tp(\lambda)$
and form the commutative family. They
are the usual Baxter $Q$-operators.

In general, one may start from the arbitrary initial states, when
$\tau_m^{(0)}$ and $\theta_m^{(0)}$ depend on $m$. (\ref{comm})
still provides the integrability of the model. But even on the
first step one gets the \emph{inhomogeneous} linear recursion for
$(\tau_m')^N$, see (\ref{linear-recursion}). Being considered with
$m\in\mathbb{Z}_M$, this linear recursion is the null eigenvector
problem for $M\times M$ matrix, constructed with the help of
$(\tau_m)^N$, $(\theta_m)^N$ and $\lambda_\ff^N$. Such equations
may be solved if the determinant of this $M\times M$ matrix is
zero. This condition defines a hyperelliptic curve $\Gamma_{M-1}$
with the genus $g=M-1$ in general position (see the next section
for the details). $(\tau_m^{(n)})^N$ and $(\theta_m^{(n)})^N$ may
be parameterized in terms of $\theta$-functions on
$\textrm{Jac}(\Gamma_{M-1})$. Our expressions (\ref{h-definition})
may be considered as the rational limit of hyperelliptic
$\theta$-functions. The most interesting feature of such
inhomogeneous models is that one may not construct at all a
commutative family of $\Q$-operators.

\section{The Baxter equation}

In this section we derive the Baxter $T-Q$ equation in the most
general operator form. We will use the well known method of the
triangulization of the auxiliary $L$-operators.

\subsection{Generic inhomogeneous chain}

Let us consider the inhomogeneous quantum chain, such that all $u_m$,
$w_m$ are in general position. Let the normalization (\ref{gauge-fix}) be
implied nevertheless. Turn to the quantum intertwining relation
(\ref{op-recursion}) and formulas (\ref{LL-darboux},
\ref{ff-recursion}, \ref{uw-prime}).

The auxiliary $L$-operator (\ref{ldst-uwkappa}) has the
degeneration point at $\lambda=\lambda_\ff$:
\begin{equation}\label{degen}
\ds\ldst_\ff(\lambda_\ff,\lambda_\ff)\;=\;
{1-\omega^{1/2}\kappa_\ff^{}\wop_\ff^{} \choose
-\omega^{1/2}\lambda_\ff^{}\uop_\ff^{-1}\wop_\ff^{}}\,\cdot\,
\biggl(\;1\;,\;-\omega^{1/2}\lambda_\ff^{-1}\uop_\ff^{}\biggr)\;.
\end{equation}
Analogously, $\ldst_\ff^{-1}(\lambda\mapsto\lambda_\ff)$ also may
be decomposed with the help of two vectors, orthogonal to ones in
(\ref{degen}). In the degeneration point quantum intertwining
relation (\ref{op-recursion}) may be rewritten in the vector
form. To do this, we will use the row vector in (\ref{degen}) and
the column vector, orthogonal to it:
\begin{equation}\label{psi-vectors}
\ds \psi^{\prime *}_{\ff,m}\;\stackrel{def}{=}\;
\biggl(1\;,\;-\omega^{1/2}\,{u_{\ff,m}\over\lambda_\ff}\,\xop_\ff\biggr)
\;\;\; \textrm{and}\;\;\; \psi^{\prime\prime}_{\ff,m}
\;\stackrel{def}{=}\;\left(\begin{array}{c}
\ds\omega^{1/2}{u_{\ff,m}\over\lambda_\ff}\xop_\ff^{}\\
\\
\ds  1
\end{array}\right)\;.
\end{equation}
Here we use the notation $\psi$ for the column vectors with
operator-valued entries, and $\psi^*$ for the similar row vectors.
The following two equations are the triangular form of
intertwining relation (\ref{op-recursion}) (the dot stands for the
matrix multiplication):
\begin{equation}\label{degen-1}
\ds\psi^{'*}_{\ff,m} \,\cdot\,
\ltoda_m(\lambda_\ff\,;u_m^{},w_m^{})\; \R_{m,\ff}^{}\;=\;
\R_{m,\ff}^{\prime}\,\;\psi^{'*}_{\ff,m+1}\;,
\end{equation}
and
\begin{equation}\label{degen-2}
\ds \ltoda_m(\lambda_\ff\,;u_m',w_m')\,\cdot\,\R_{m,\ff}
\;\psi^{\prime\prime}_{\ff,m+1} \;=\;
\psi^{\prime\prime}_{\ff,m}\;\R_{m,\ff}^{\prime\prime}\;.
\end{equation}
Here
\begin{equation}\label{R-12}
\ds\begin{array}{l} \ds \R_{m,\ff}^{\prime}\;=\; {u_m\over
u_{\ff,m+1}}\;\;\xop_m^{}\;\R_{m,\ff}\;\xop_\ff^{-1}\;,\\
\\
\ds \R_{m,\ff}^{\prime\prime}\;=\;
\omeg\,{u_{\ff,m+1}\,w_{m}\over\lambda_\ff}\;\;
\zop_m^{}\;\R_{m,\ff}\;\xop_\ff^{}\;,
\end{array}\end{equation}
and the recursion relations (\ref{ff-recursion}, \ref{uw-prime}) are
implied. Equations (\ref{degen-1}, \ref{degen-2}, \ref{R-12})
can be verified by the direct calculations in the component
form.

Let further $\Q^{\prime}_\ff$ and $\Q^{\prime\prime}_\ff$ be the
traces of the monodromies of $\R_{m,\ff}^{\prime}$ and
$\R_{m,\ff}^{\prime\prime}$ similar to (\ref{R-monodromy}) and we imply the
$\mathbb{Z}_M$ boundary conditions for the chain and recursion
(\ref{ff-recursion}). For two vectors $\psi^{\prime
*}_{\ff,1}=\psi^{\prime *}_{\ff,M+1}$ and
$\psi''_{\ff,1}=\psi''_{\ff,M+1}$ let $\psi'_{\ff,1}$ and
$\psi^{\prime\prime *}_{\ff,1}$ be their dual vectors
\begin{equation}
\ds \psi^{\prime *}\cdot\psi'\;=\; \psi^{\prime\prime
*}\cdot\psi''\;=\;1\;,\;\;\; \psi^{\prime *}\cdot\psi''\;=\;
\psi^{\prime\prime *}\cdot\psi'\;=\;0\;,
\end{equation}
such that the $2\times 2$ unity matrix may be decomposed as follows
\begin{equation}
\ds \psi'\;\psi^{\prime *}\,+\,\psi''\;\psi^{\prime\prime
*}\;=\;1\;.
\end{equation}
This decomposition allows one to calculate
\begin{equation}\label{be-corollary}
\ds\begin{array}{l} \ds \tp(\lambda_\ff)\,\Q_{\ff}\;=\;
\textrm{tr}_{\mathbb{C}^2,\ff}\, \biggl(
(\psi'_{\ff,1}\;\psi^{\prime *}_{\ff,1} \,+\,
\psi''_{\ff,1}\;\psi^{\prime\prime *}_{\ff,1})
\,\cdot\,\montoda(\lambda_\ff)\,\widehat{\Q}_\ff\biggr)\\
\\
\ds =\;\textrm{tr}_{\mathbb{C}^2,\ff}\, \biggl(
\psi'_{\ff,1}\psi^{\prime *}_{\ff,1}\cdot
\montoda(\lambda_\ff)\,\widehat{\Q}_\ff\,+\,
\montoda(\lambda_\ff)\,\widehat{\Q}_\ff\cdot
\psi''_{\ff,M+1}\psi^{\prime\prime *}_{\ff,M+1}\biggr)\\
\\
\ds =\; \textrm{tr}_{\ff}\,\biggl(\psi^{\prime
*}_{\ff,1}\cdot\widehat{\Q}_\ff^{}\psi'_{\ff,1} \,+\,
\psi^{\prime\prime *}_{\ff,1}\cdot \widehat{\Q}_\ff^{}
\psi^{\prime\prime}_{\ff,1}\biggr)\;=\;\Q_\ff'\,+\,\Q_\ff''\;.
\end{array}
\end{equation}
The last equality here is obtained using recursions
(\ref{degen-1}) and (\ref{degen-2}). Eq. (\ref{be-corollary}) gives
the Baxter equation in its operator form
\begin{equation}\label{BE-operator-12}
\ds \tp(\lambda_\ff)\;\Q_\ff\;=\; \Q_\ff\;\tp'(\lambda_\ff)\;=\;
\Q^{\prime}_\ff\;+\;\Q^{\prime\prime}_\ff\;,
\end{equation}
where $\tp(\lambda)$ is the initial transfer matrix with the
parameters $u_m,w_m$, and $\tp'(\lambda)$ is the transfer matrix
with the parameters $u_m',w_m'$. Using (\ref{R-12}) and the first
relation of (\ref{finite-mapping}), one may obtain
\begin{equation}
\ds \Q^{\prime}_\ff=\prod_m{u_m\over
u_{\ff,m}}\;\mathbf{X}\Q_\ff\mathbf{X}^{-1}\;,\;\;\;
\Q^{\prime\prime}=\prod_{m}\omega^{1/2}{u_{\ff,m}w_m\over\lambda_\ff}\;
\mathbf{Z}\mathbf{Q}_\ff\mathbf{X}\;,
\end{equation}
where $\mathbf{X}$ and $\mathbf{Z}$ are given by
\begin{equation}\label{XZ}
\ds\mathbf{X}=\prod_{m=1}^M\,\xop_m\;,\;\;\;
\mathbf{Z}=\prod_{m=1}^M\zop_m\;.
\end{equation}
Note that in the spectral decomposition of $\tp(\lambda)$ and
$\tp'(\lambda)$ there exists operator $\mathbf{Y}$ (\ref{Y}).
$\Q^{\prime}_\ff$ may be rewritten as
\begin{equation}
\ds \Q^{\prime\prime}_\ff\;=\;\left(\prod_{m=1}^M\,-\,{u_{\ff,m}
w_m\over\lambda_\ff}\right)\;
\mathbf{Y}\;\mathbf{X}^{-1}\,\Q_\ff\,\mathbf{X}\;.
\end{equation}
Using (\ref{Q-elements}), one may check
$\mathbf{Y}\Q_\ff=\Q_\ff\mathbf{Y}$ (the matrix elements should be
taken). Let now (see (\ref{gauge-fix}))
\begin{equation}\label{mu-def}
\ds \prod_{m=1}^M\,{u_{\ff,m}\over
u_m}\;=\;\mu_\ff\;,\;\;\;
\prod_{m=1}^M\,u_{\ff,m}w_m\;=\;\omega^{-M/2}\,\mu_\ff\;.
\end{equation}
The Baxter equation in the operator form becomes
\begin{equation}\label{BE-op}
\ds \tp(\lambda_\ff)\;\Q_\ff\;=\; \Q_\ff\;\tp'(\lambda_\ff)\;=\;
\mu_\ff^{-1}\,\mathbf{X}\,\Q_\ff\,\mathbf{X}^{-1}\,+\,
{\mu_\ff\over \left(-\omega^{1/2}\lambda_\ff\right)^M}\,
\mathbf{Y}\, \mathbf{X}^{-1}\, \Q_\ff\, \mathbf{X}\;.
\end{equation}
The natural question arises: what is $\mu_\ff$? To answer it,
consider (\ref{LL-darboux}) in the degeneration point
$\lambda=\lambda_\ff$, applied to the classical monodromy matrix
$\Montoda$, eqs. (\ref{Mon-toda}) and (\ref{LT-darboux}). Using
the formulas like (\ref{degen-1}) and (\ref{degen-2}) for the
classical counterpart, one obtains
\begin{equation}
\ds \begin{array} {l} \ds \left(1\,,
\;{u_{\ff,1}^N\over\lambda_\ff^N}\right) \,\cdot\,
\Montoda(\lambda_\ff^N) \;=\; \prod_{m=1}^M{u_m^N\over
u_{\ff,m+1}^N}\;\cdot\;
\left(1\,,\;{u_{\ff,M+1}^N\over\lambda_\ff^N}\right)\;,\\
\\
\ds \Montoda(\lambda_\ff^N)\,\cdot\,\left(\begin{array}{c} \ds
-{u_{\ff,M+1}^N\over\lambda_\ff^N} \\ \ds 1\end{array}\right)
\;=\; \left(\begin{array}{c} \ds -{u_{\ff,1}^N\over\lambda_\ff^N}
\\ \ds 1\end{array}\right)\;\prod_{m=1}^M\,\left(-{u_{\ff,m+1}^N
w_m^N\over \lambda_\ff^N}\right)\;.
\end{array}
\end{equation}
When the $\mathbb{Z}_M$ boundary condition $u_{\ff,M+1}=u_{\ff,1}$
is imposed, one may see that both $\mu_\ff^{-N}$ and
$\mu_\ff^N/(\lambda_\ff^N)^M$ are the eigenvalues of the classical
monodromy matrix, i.e.
\begin{equation}\label{classical-curve}
\ds J(\lambda_\ff^N,\mu_\ff^N)\;\stackrel{def}{=}\;
\det\left(\mu_\ff^{-N}\,-\,\Montoda(\lambda^N_\ff)\right)\;=\;0
\end{equation}
or
\begin{equation}
\ds A(\lambda_\ff^N)+D(\lambda_\ff^N) \;=\;
{1\over\mu_\ff^N}\,+\,{\mu_\ff^N\over\lambda_\ff^{NM}}\;.
\end{equation}
Hence $(\lambda_\ff^N,\mu_\ff^N)$ is the point of the genus $g=M-1$
hyperellitic curve $\Gamma_{M-1}$, defined by
(\ref{classical-curve}). As it was mentioned in the section
(\ref{sec-inhomo}) on page \pageref{sec-inhomo}, all $u_m^N$,
$w_m^N$, $u_m^{\prime N}$ and $w_m^{\prime N}$ may be
parameterized in the terms of $\theta$-functions on
$\textrm{Jac}(\Gamma_{M-1})$.

Operator $\Q_\ff$ makes the isospectrality transformation
(\ref{BE-operator-12}) of the initial inhomogeneous quantum chain.
So it may be decomposed similar to  (\ref{K-decomposition}). If
$|\Psi_t\rangle$ is the complete basis of the eigenvectors of
$\tp(\lambda)$ then
\begin{equation}\label{Q-decomposition}
\ds \Q_\ff\;=\;\sum_{t}\;|\Psi_t^{}\rangle\; q_{t,\ff}\;
\langle\Psi'_t|\;.
\end{equation}

Using further the explicit formula (\ref{Q-elements}) for the
matrix elements of $\Q_\ff$ and the properties (\ref{w-simp}) of
$w_p$-functions, one may calculate (taking the matrix elements)
\begin{equation}\label{Q-X-prop}
\ds \mathbf{X}\, \Q_\ff \,\mathbf{X}^{-1}\;=\;
\Q_{\ff}[y_{1,m}\mapsto \omega^{-1}y_{1,m}]\;.
\end{equation}
and besides
\begin{equation}\label{Q-Y-prop}
\ds\begin{array}{l} \ds
\mathbf{Y}\,\Q_{\ff}\;=\;\Q_\ff\,\mathbf{Y}\;=\\
\\
\ds \prod_{m\in\mathbb{Z}_M}\,\left( -\omega^{-1/2}\,{y_{1,m}\over
y_{3,m}}\, {1-\omega x_{3,m}\over 1-\omega
x_{1,m}}\right)\,\cdot\, \Q_{\ff}[x_{1,m},y_{1,m},x_{3,m} \mapsto
\omega x_{1,m} \omega^{-1} y_{1,m}, \omega x_{3,m}]\;.
\end{array}
\end{equation}
The square braces in the right
hand sides of these two equations show the changes of the
Fermat curve points. The rest $x_{j,m}$ and $y_{j,m}$ are the same
in the left and right hand sides.

A glance to the parametrization (\ref{xy-explicit}) shows that the
simultaneous change of the phases of all, say, $y_{1,m}$ is
equivalent to the change of $\lambda_\ff$ while $\delta_\ff$,
$\mu_\ff$ and $w_\ff$ remain unchanged. Using (\ref{Q-X-prop}),
we may conclude
\begin{equation}\label{XQX}
\ds \mathbf{X}\Q_\ff\mathbf{X}^{-1} \;=\;
\sum_{t}\;|\Psi_t^{}\rangle\;
q_{t,\ff}(\lambda_\ff\mapsto\omega^{-1}\lambda_\ff)\;
\langle\Psi'_t|\;.
\end{equation}
Besides,  both $|\Psi_t^{}\rangle$ and
$\langle\Psi'_t|$ are the eigenvectors of $\mathbf{Y}$,
$Y\;|\Psi_t^{}\rangle\;=\;|\Psi_t^{}\rangle\;\omega^\gamma$.
The operator equation (\ref{BE-op}) may be rewritten as the functional
equation
\begin{equation}\label{BE}
\ds t(\lambda)\;q_t(\lambda,\mu)\;=\;
\mu^{-1}\,q_t(\omega^{-1}\lambda,\mu)\;+\;
{\mu\over\left(-\omega^{1/2}\lambda\right)^N}\,
\omega^\gamma\,q_t(\omega\lambda,\mu)\;,
\end{equation}
where $(\lambda^N,\mu^N)\in\Gamma_{M-1}$, eq.
(\ref{classical-curve}), and in the decomposition
$t(\lambda)=\sum_{m=0}^M\lambda^{-m}t_m$,
$t_M=(-\kappa)^M\omega^\gamma$.

Note that equation (\ref{BE}) does not follow from (\ref{BE-op}) directly.
In general, coefficients
$q_{t,\ff}=\langle\Psi_t^{}|\Q_\ff|\Psi_t'\rangle$
in (\ref{BE-op}) may have the following
structure
\begin{equation}\label{Q-stupid}
\ds q_{t\,\ff}\;=\;q_t(\lambda_\ff,\mu_\ff;\{u_m^{},w_m^{}\}_{m=1}^M,
\{u_m',w_m'\}_{m=1}^M)\;.
\end{equation}
Since the matrix elements of $\Q_\ff$ are the rational functions
of $\lambda_\ff$ and of $\mu_\ff$, $q_t$ is a rational function of
$\lambda_\ff$ and $\mu_\ff$. Therefore due to (\ref{XQX}) the
matrix elements of (\ref{BE-op}) between $\langle\Psi_t^{}|$ and
$|\Psi_t'\rangle$ are the system of $N$ linear equations for
$q_t(\omega^k\lambda_\ff,\mu_\ff;...)$, $k\in\mathbb{Z}_N$. Since
$(\lambda_\ff^N,\mu_\ff^N)\in\Gamma_{M-1}$, the rank of this
linear system is equal to $N-1$ for generic $\lambda_\ff$ (see
\cite{Sklyanin-fun} for the details). Hence the linear system has
a unique solution in the class of the rational functions of
$\lambda_\ff$, $\mu_\ff$ up to a multiplier, depending on
$\mu_\ff$ in general. We will show further that such multiplier
may be fixed by an extra relation provided by (\ref{Q-Y-prop})).

Actually, there are two spectral curves in our considerations.
The first one is the classical spectral curve $\Gamma_{M-1}$ with the point
$(\lambda_\ff^N,\mu_\ff^N)$ (\ref{classical-curve}). But the point
$(\lambda_\ff,\mu_\ff)$ belongs to the quantum curve,
which is $N^2$-sheeted covering of the classical one.

The considerations above prove the following
\begin{prop}\label{prop-prop}
For any isospectral transformation $\{u_m,w_m\}_{m=1}^M$ $\mapsto$
$\{u_m',w_m'\}_{m=1}^M$ the coefficients $q_t$ of the
corresponding modified $Q$-operator, decomposed as in
(\ref{Q-decomposition}), depend only on the moduli of the spectral
curve and on the point on the spectral curve:
\begin{equation}
\ds q_{t,\phi}\;=\;q_t(\lambda_\ff,\mu_\ff)\;,
\end{equation}
and do not depend on ``classical times'' (i.e. a point on the
jacobian), entering to a parameterization of $u_m,w_m,u_m',w_m'$.
In particular, for the homogeneous initial state, $q_t$ does not depend
on the amplitudes of the solitonic states involved.
\end{prop}
This proposition clarifies the derivation of (\ref{BE}) from (\ref{BE-op}).

\subsection{Baxter equation for the homogeneous chain}

All the formulas from the previous subsection are valid for the
homogeneous initial state (\ref{uw-homo}). Arbitrary value of
$\lambda_\ff$ provides the trivial action of the functional
counterpart $\mathcal{Q}_\ff$ of $\Q_\ff$, see discussion after
proposition \ref{prop-inhomo} on page \pageref{prop-inhomo}.
Such $\Q$-operators commute with $\tp(\lambda)$ and form the
commutative family.
Explicit form of this $\Q$-operator is given by
(\ref{Q-elements}) with the homogeneous parameterization
$p_{1,m}=p_1$, $p_{2,m}=p_2$, $p_{3,m}=p_3$,
\begin{equation}\label{xy-homo}
\ds\begin{array}{l} \ds
x_1=\omega^{-1/2}{\delta_\ff\over\kappa}\;,\;\;
x_2=\omega^{-1/2}\kappa\;,\;\;x_3=\omega^{-1}\delta_\ff\;,\\
\\
\ds {y_3\over y_2}= w_\ff\;,\;\; {y_1\over
y_3}=\omega^{1/2}\,{\lambda_\ff\over\kappa\delta_\ff}\;,
\end{array}
\end{equation}
where $\delta_\ff$, $\lambda_\ff$ and $w_\ff$ are given by
(\ref{phi-parametrization}, \ref{delta-Delta}, \ref{w-phi-m})
with generic value of $\phi$. Due to (\ref{xy-homo})
\begin{equation}
\ds \mathbf{Q}\;=\;\mathbf{Q}(\lambda_\ff,\delta_\ff)\;.
\end{equation}
Several properties of this $\Q$-operator may be derived
with the help of the matrix elements of (\ref{Q-elements}) and
(\ref{xy-homo}), and with the help of (\ref{w-simp}). In particular,
(\ref{Q-X-prop}) and (\ref{Q-Y-prop}) become
\begin{equation}\label{Q-prop}
\ds
\begin{array}{l}
\ds
\mathbf{X}\mathbf{Q}(\lambda_\ff,\delta_\ff)\mathbf{X}^{-1}\;=\;
\mathbf{Q}(\omega^{-1}\lambda_\ff,\delta_\ff)\;,\\
\\
\ds \mathbf{Y}\mathbf{Q}(\lambda_\ff,\delta_{\ff}) \;=\;
\mathbf{Q}(\lambda_\ff,\delta_\ff)\mathbf{Y}\;=\;
\left(-{\lambda_\ff\over\delta_\ff}\,
{1-\delta_\ff\over\kappa-\omega^{1/2}\delta_\ff}\right)^M\,
\mathbf{Q}(\lambda_\ff,\omega\delta_\ff)\;.
\end{array}
\end{equation}
The second equation is the consequence of (\ref{Q-Y-prop}) since
\begin{equation}
\ds x_1,y_1,x_3\mapsto\omega x_1, \omega^{-1} y_1, \omega x_3\;\;
\Leftrightarrow\;\; \delta_\ff\mapsto \omega\delta_\ff\;.
\end{equation}
Using (\ref{Q-prop}) and (\ref{BE-op}), we
obtain
\begin{equation}\label{BE-operator}
\ds\tp(\lambda_\ff)\Q(\lambda_\ff)=
\delta_\ff^{-M}\Q(\omega^{-1}\lambda_\ff)\,+\,
\left(-\omega^{-1/2}{\delta_\ff\over\lambda_\ff}\right)^M
\mathbf{Y}\Q(\omega\lambda_\ff)\;,
\end{equation}
where $\delta_\ff$-argument of all $\Q$-s remains unchanged. Recall that
$\tp$ and $\Q$ may be diagonalized simultaneously. Comparing
(\ref{BE-operator}) and (\ref{BE-op}), we conclude $\mu_\ff\equiv\delta_\ff^M$.
Further we
will omit the subscribe $\phi$.  Let $t(\lambda)$ and
$q_t(\lambda,\delta)$ be the eigenvalues of $\tp(\lambda)$ and $\Q(\lambda)$
for the same eigenvector, then (\ref{BE-operator}) provides the functional
equation
\begin{equation}\label{simple-BE}
\ds t(\lambda) q_t(\lambda,\delta)= \delta^{-M}
q_t(\omega^{-1}\lambda,\delta)\,+\,
\left(-\omega^{-1/2}{\delta\over\lambda}\right)^M\omega^\gamma
q_t(\omega\lambda,\delta)\;.
\end{equation}
Here $q_t$
is a meromorphic function on the Baxter curve (see the set of
definitions (\ref{Delta-curve}, \ref{delta-Delta}) etc.):
\begin{equation}\label{Baxter-curve}
\ds (\delta^{-1},\delta^{*-1})\in\Gamma_B\;\;\Leftrightarrow\;\;
{1\over\delta^{N}} \,+\, {1\over\delta^{*N}} \;=\; 1 \,-\,
\kappa^N\,{1\over\delta^N\delta^{*N}}\;,\;\;\;
\lambda^{}=\delta^{}\delta^*\;.
\end{equation}
More detailed investigation of (\ref{simple-BE}) in the spirit of
Proposition \ref{prop-prop} on page \pageref{prop-prop} shows that
for generic $\lambda$ any solution of (\ref{simple-BE}) such that
$t(\lambda)$ is a polynomial and $q_t(\lambda,\delta)$ is a
meromorphic function on the curve (\ref{Baxter-curve}), gives the
eigenvalue of $\tp(\lambda)$ and $\mathbf{Q}(\lambda)$. The second
equation of (\ref{Q-prop}) provides the equation, fixing the
$\delta$ ambiguity of the solution of the Baxter equation
(\ref{simple-BE}):
\begin{equation}
\ds \omega^\gamma\,q_t(\lambda,\delta)\;=\;
\left(-{\lambda\over\delta}\,
{1-\delta\over\kappa-\omega^{1/2}\delta}\right)^M \;
q_t(\lambda,\omega\delta)\;.
\end{equation}

\subsection{Evolution operator}

$\mathcal{Q}_\ff$-transformation and its quantum counterpart
$\Q_\ff$ may be interpreted as a kind of evolution,
depending strongly on their spectral parameter $\lambda_\ff$.
Evolution in more usual sense may be obtained in the limit when
$\lambda_\ff^{-1}\mapsto 0$. Such a phenomenon, when a
``physical'' evolution operator is $Q$-operator in the singular
point, appeared for example in the quantum Liouville model \cite{fkv}.

Evolution transformation $\mathfrak{E}$ for the quantum
relativistic Toda chain may be produced by non-local similarity
transformation of the quantum $L$-operators (\ref{l-toda}):
\begin{equation}\label{A-similarity}
\ds \mathfrak{E}(\ell_m)(\lambda) \;=\;
\mathbf{A}_m(\lambda)^{-1}\,\ell_m(\lambda)\,
\mathbf{A}_{m+1}(\lambda)\;,
\end{equation}
where $2\times 2$ matrix $\mathbf{A}_m$ has the operator-valued
entries:
\begin{equation}\label{A}
\ds\mathbf{A}_m(\lambda)\;=\; \left(\begin{array}{ll} \ds 1 & \ds
-\lambda^{-1}\wop_{m-1}^{-1}\\
&\\
\ds \wop_m^{} & -\omega^{-1/2}\kappa\lambda^{-1}
\end{array}\right)\;.
\end{equation}
The similarity transformations do not change the
transfer matrix of the model with the periodical boundary
conditions. The explicit form of $\mathfrak{E}$-transformation may
be obtained from (\ref{A-similarity}):
\begin{equation}\label{E-hom}
\ds\left\{\begin{array}{l} \ds \mathfrak{E}(\uop_m^{})\;=\;
\omega^{-1/2}\wop_m^{-1}\;,\\
\\
\ds \mathfrak{E}(\wop_m^{})\;=\;
(1-\omega^{-1/2}\kappa\wop_m^{-1}\wop_{m-1}^{})^{-1}
\omega^{1/2}\wop_{m-1}^{}\uop_m^{}\wop_{m+1}^{}
(1-\omega^{-1/2}\kappa\wop_{m+1}^{-1}\wop_m^{})\;.
\end{array}\right.
\end{equation}
One may check directly that $\mathfrak{E}$-transformation is the
canonical one. As usual, at the  root of unity we have to separate
\begin{equation}
\ds
\mathfrak{E}(f)\;=\;\mathbf{E}\,\mathcal{E}(f)\,\mathbf{E}^{-1}\;,
\end{equation}
where $\mathbf{E}$ is the finite dimensional operator, and
$\mathcal{E}$ acts on $N$th powers of the Weyl elements
$\uop_m^N$, $\wop_m^N$ as follows
\begin{equation}\label{E-func}
\ds\left\{\begin{array}{l} \ds \mathcal{E}(u_m^{N})\;=\;
-\,{1\over w_m^{N}}\;,\\
\\
\ds \mathcal{E}(w_m^{N})\;=\; -\, u_m^N w_m^N w_{m-1}^N
{w_{m+1}^N\,+\,\kappa^N w_m^N\over w_m^N\,+\,\kappa^N
w_{m-1}^N}\;.
\end{array}\right.
\end{equation}
For the given set of $\{u_m,w_m\}_{m=1}^M$ parameterized in the
terms of $\tau_m$ and $\theta_m$ (\ref{uw-inhomo}), denote
\begin{equation}
\ds
\tau_m^{}\;=\;\tau_{0,m}^{}\;,\;\;\;\theta_m^{}\;=\;\tau_{1,m}^{}\;,
\end{equation}
where we have introduced the extra auxiliary subscript of
$\tau$-function. The first equation in (\ref{E-func}) yields
\begin{equation}
\ds\mathcal{E}(\tau_{l,m}^{})\;=\;\tau_{l+1,m}^{}\;,
\end{equation}
and the second equation in (\ref{E-func}) gives the
following ``equation of motion''
\begin{equation}\label{Hirota}
\ds \tau_{l,m+1}^N\tau_{l,m-1}^N\,+\, \kappa^N
(\tau_{l,m}^N)^2\;=\; (1+\kappa^N) \tau_{l+1,m}^N\tau_{l-1,m}^N\;.
\end{equation}

Let $\tau_{0,m}^N=\tau_{m}^N$ is given by the $n$-solitonic
expression  (\ref{tau-and-theta}). Then the complete solution
of (\ref{Hirota}) is given by
\begin{equation}\label{ev-solution}
\ds \left(\tau_{l,m}^{(n)}\right)^N\;=\; h^{(n)}(\{f_k \EXP^{2 i m
\phi_k+2 i l \psi_k}\}_{k=1}^n)\;,
\end{equation}
where
\begin{equation}
\ds\EXP^{2i\psi_k} \;=\; s_k(1) \;=\;
{\Delta_k^*\over\Delta_k^{}}\,{\Delta_k^{}-1\over\Delta_k^*-1}\;,
\end{equation}
see (\ref{h-definition},\ref{s-function}) for the notations.
The substitution (\ref{ev-solution}) makes
(\ref{Hirota}) the identity (see the Appendix).

Comparing (\ref{ev-solution}) with the definition of $h^{(g)}$ and
$\tau_m^{(g)}$ (\ref{tau-and-theta},\ref{h-definition}),
we see that in fact
\begin{equation}
\ds\tau^{(n)}_{l,m}\;=\;\tau^{(n+l)}_m\;,
\end{equation}
where the initial set of $\{\phi_k,f_k\}_{k=1}^n$ for $h^{(n)}$
and $\tau^{(n)}_m$ is enlarged formally by
$\{\phi_k',f_k'\}_{k=n+1}^{n+l}$ with $f_k'\equiv 0$ and all
$\phi_k'$ equal and obeying formally
\begin{equation}
\ds \delta_{\phi_k'}^{*N}\;=\;1\;\;\;\Rightarrow\;\;\;
\delta_{\phi_k'}^{-1}\,=\,\lambda_{\phi_k'}^{-1}\;=\;0\;.
\end{equation}
This establishes the relation between the evolution (\ref{E-hom},\ref{E-func})
and $\mathcal{Q}_\ff$, $\Q_\ff$.
In particular, the finite dimensional counterpart $\mathbf{E}$ of the
evolution operator $\mathfrak{E}$ may be obtained in the limit
\begin{equation}
\ds \mathbf{E}\;=\;\lim_{\delta=\infty,\;\;\delta^*=1}\;
\Q(\delta,\delta^*)\;.
\end{equation}
Eq. (\ref{Q-elements}) gives the explicit form of the matrix elements of
$\mathbf{E}$ on the $l$-th step of the evolution
\begin{equation}\label{E-elements}
\ds \langle\alpha|\mathbf{E}_l|\beta\rangle\;=\; \prod_{m=1}^M\,
\omega^{(\alpha_m-\beta_{m+1})\beta_m}\,
w_{p_{2,l,m}}(\beta_{m+1}-\beta_m)\;,
\end{equation}
where
\begin{equation}
\ds  x_{2,l,m}\;=\;\omega^{-1/2}\kappa\,
{\tau_{l,m}^2\over\tau_{l,m-1}^{}\tau_{l,m+1}^{}}\;,
\;\;y_{2,l,m}\;=\; \kappa'\,{\tau_{l-1,m}^{}\tau_{l+1,m}^{}
\over\tau_{l,m-1}^{}\tau_{l,m+1}^{}}
\end{equation}
with
\begin{equation}
\ds\kappa^{\prime N}\;=\;1\,+\,\kappa^N\;.
\end{equation}

In particular, $N$-bilinear relation (\ref{Hirota}) allows one to
parameterize the Fermat relation for the coordinates of $p_{j,m}$. Let
\begin{equation}\label{theta-tilde}
\ds\overline{\theta}_m^{}\;=\;\tau_{-1,m}^{}\;.
\end{equation}
Then
\begin{equation}
\ds\begin{array}{l}
\ds y_{1,m}^{(n)}\;=\; \omega^{1/2}{\lambda_n w_{\phi_n} \kappa'
\over \kappa \delta_n} {\overline{\theta}_m^{(n)}
\theta_{m-1}^{(n-1)} \over \tau_m^{(n)} \tau_{m-1}^{(n-1)}} \;,\\
\\
\ds
y^{(n)}_{2,m}\;=\;\kappa'\,
{\overline{\theta}_m^{(n)}\theta_m^{(n)}
\over\tau_{m-1}^{(n)}\tau_{m+1}^{(n)}}\;,\;\;\; y_{3,m}^{(n)}\;=\;
w_{\ff_n}\kappa'\, {\overline{\theta}_m^{(n)} \theta_{m}^{(n-1)}
\over \tau_{m+1}^{(n)}\tau_{m-1}^{(n-1)}}\;.
\end{array}
\end{equation}

\section{Quantum separation of variables}

Turn to the last subject of this paper -- the quantum separation
of variables, or functional Bethe ansatz.

In the previous sections we have established the following.
Let $\tp(\lambda)\equiv\tp^{(0)}(\lambda)$ be the transfer matrix
of the homogeneous quantum chain of the length $M$.
There exists the family of isospectral to
$\tp(\lambda)$ transfer matrices $\tp^{(M-1)}(\lambda)$, parameterized
by $(M-1)$ complex variables $f_1,...,f_{M-1}$ -- the amplitudes
of Hirota-type solitons in the minimal set (\ref{minimal-set}).
The isospectrality means that for any generic values of
$f_1,...,f_{M-1}$ $\tp^{(M-1)}(\lambda)$ has the same eigenvalues
as the initial $\tp(\lambda)$. Therefore only the eigenvectors
$|\Psi_t^{(M-1)}\rangle$ depend on $f_1,...,f_{M-1}$. We may
choose the values of $f_1,...,f_{M-1}$ in an appropriate
way, such that $|\Psi_t^{(M-1)}\rangle$ would become more simple
then the initial unknown $|\Psi_t\rangle$. Since the similarity
matrix $\mathbf{K}^{(M-1)}$ is known explicitly, any exact
information about $|\Psi^{(M-1)}\rangle$ solves
the eigenstate problem for the initial homogeneous quantum
relativistic Toda chain.

\subsection{Null subspaces of $\mathbf{b}(\lambda)$}

Now we have to understand what special values of $f_1,...,f_{M-1}$
are the most useful ones. To do this, we should turn back to the
degenerated $\ldst_\ff$-operator
and write out equations (\ref{degen-1}) and (\ref{degen-2})
for the monodromies of $\ltoda$ and $\R$
\begin{equation}
\ds\begin{array}{l} \psi^{\prime
*}_{\ff,1}\,\cdot\,\montoda(\lambda_\ff)\;\widehat{\Q}_\ff^{}\;=\;
\widehat{\Q}_\ff'\;\psi^{\prime *}_{\ff,M+1}\;,\\
\\
\ds \montoda(\lambda_\ff)\;\widehat{\Q}_{\ff}^{}\,\cdot\,
\psi''_{\ff,M+1}\;=\;\psi''_{\ff,1}\;\widehat{\Q}_{\ff}''\;.
\end{array}
\end{equation}
In components,
\begin{equation}\label{Q-mon-1}
\ds\begin{array}{l} \ds\left(\mathbf{a}(\lambda_\ff)\,-\,
\omega^{1/2} {u_{\ff,1}\over\lambda_\ff}\xop_\ff^{}
\mathbf{c}(\lambda_\ff)\right)\,\widehat{\Q}_\ff^{}\;=\;
\widehat{\Q}_\ff^\prime\;,\\
\\
\ds \left(\mathbf{b}(\lambda_\ff)\,-\, \omega^{1/2}
{u_{\ff,1}\over\lambda_\ff}\xop_\ff^{}
\mathbf{d}(\lambda_\ff)\right)\,\widehat{\Q}_\ff^{}\;=\;
-\omega^{1/2}{u_{\ff,M+1}\over\lambda_\ff}\,
\widehat{\Q}_\ff^\prime\,\xop_\ff^{}\;,
\end{array}
\end{equation}
and
\begin{equation}\label{Q-mon-2}
\ds\begin{array}{l} \ds
\mathbf{a}(\lambda_\ff)\widehat{\Q}_\ff^{}\,
\omega^{1/2}{u_{\ff,M+1}\over\lambda_\ff}\,\xop_\ff^{}\,+\,
\mathbf{b}(\lambda_\ff)\,\widehat{\Q}_\ff^{}\;=\;
\omega^{1/2}\,{u_{\ff,1}\over\lambda_\ff}\,\xop_\ff^{}
\widehat{\Q}_\ff^{\prime\prime}\;,\\
\\
\ds \mathbf{c}(\lambda_\ff)\widehat{\Q}_\ff^{}\,
\omega^{1/2}{u_{\ff,M+1}\over\lambda_\ff}\,\xop_\ff^{}\,+\,
\mathbf{d}(\lambda_\ff)\,\widehat{\Q}_\ff^{}\;=\;
\widehat{\Q}_\ff^{\prime\prime}\;.
\end{array}
\end{equation}
Using relation $\R_{m,\ff}\xop_m=\xop_\ff\R_{m,\ff}$ (see the
(\ref{finite-mapping})) one may obtain from (\ref{R-12})
\begin{equation}
\ds\widehat{\Q}_\ff^\prime\;=\; \left(\prod_{m=1}^M\,{u_m\over
u_{\ff,m+1}}\right)\, \mathbf{X}\, \widehat{\Q}_\ff^{}\,
\mathbf{X}^{-1}\,\xop_1^{}\xop_\ff^{-1}\;,
\end{equation}
and
\begin{equation}
\ds\widehat{\Q}_\ff^{\prime\prime}\;=\;
\left(\prod_{m=1}^M\,\omega^{1/2}{u_{\ff,m+1} w_m\over
\lambda_\ff}\right)\, \mathbf{Z}\, \widehat{\Q}_\ff^{}\,
\mathbf{X}^{}\,\xop_1^{-1}\xop_\ff^{}\;.
\end{equation}

Let the initial chain is the homogeneous one.
If $\phi\in\Set$, then
\begin{equation}
\ds u_{\ff,m}^N\;=\;u_{\ff,M+m}^N\;=\;(-)^{N-1}\delta_\ff^N\;
{1\,-\,f_\ff\EXP^{2i\phi(m-1)}\over 1\,-\,f_\ff\EXP^{2i\phi m}}\;.
\end{equation}
Note,
\begin{equation}\label{produphim}
\ds\prod_{m\in Z_M} u_{\ff,m}\;=\;(-\omega^{-1/2}\delta_\ff)^M\;
\end{equation}
does not depend on $f_\ff$. Consider the limit when $f_\ff\mapsto
1$. Due to (\ref{produphim}) $\widehat{\Q}_\ff$,
$\widehat{\Q}_\ff'$ and $\widehat{\Q}_\ff''$ are regular, but
\begin{equation}
\ds u_{\ff,1}\;=\;u_{\ff,M+1}\;=\;0\;,
\end{equation}
and (\ref{Q-mon-1},\ref{Q-mon-2}) give
\begin{equation}
\ds \mathbf{a}(\lambda_\ff)\, \widetilde{\Q}_\ff^{} \;=\;
\widetilde{\Q}_\ff'\;,\;\;\; \mathbf{d}(\lambda_\ff)\,
\widetilde{\Q}_\ff^{} \;=\; \widetilde{\Q}_\ff''\;,\;\;\;
\mathbf{b}(\lambda_\ff)\, \widetilde{\Q}_\ff^{} \;=\; 0\;.
\end{equation}
The last expression is important. Taking the trace over $\ff$-th
space, one obtains
\begin{equation}
\ds
\mathbf{b}(\lambda_\ff)\;\Q_\ff\;=\;0\;\;\;\textrm{if}\;\;f_\ff=1\;.
\end{equation}
Recall, the phase of $\lambda_\ff$ is the parameters of $\Q_\ff$.

Turn now to the operator $\mathbf{K}^{(M-1)}$ and consider the
limit when all $f_n\mapsto 1$, $n=1,...,M-1$. Each $\Q_{\phi_n}$
in (\ref{K-operator}) depends essentially on the phase of $\lambda_{\phi_n}$,
$n=1,...,M-1$. Since $\mathbf{K}^{(M-1)}$ is the symmetrical
operator with respect to any permutation of
$\{(f_n,\phi_n)\}_{n=1}^{M-1}$, see Proposition
\ref{prop-ordering} on page \pageref{prop-ordering}, the
following proposition holds
\begin{prop}\label{prop-null}
Let $\lambda_{\phi_n}$ be the spectral parameter of $\Q_{\phi_n}$ in the
decomposition (\ref{K-operator}) of the minimal
$\mathbf{K}^{(M-1)}$ (\ref{minimal-set}). Then
\begin{equation}\label{K-null}
\ds \mathbf{b}(\lambda_{\ff_n})\;\mathbf{K}^{(M-1)}\;=\;0
\;\;\;n=1,...,M-1\;\;\; \textrm{if all}\;\;f_n\mapsto 1\;.
\end{equation}
\end{prop}
The structure of eigenvectors of $\mathbf{b}(\lambda)$ was
described by eq. (\ref{b-eigen}). Comparing (\ref{b-eigen}) and
(\ref{K-null}), and taking into account
$\mathbf{Y}\Q_\ff\;=\;\Q_\ff\mathbf{Y}$ for any $\Q_\ff$, we may
conclude that in the limit $f_k\mapsto 1$ $\forall k$
\begin{equation}\label{K-separation}
\ds \mathbf{K}^{(M-1)}({\{\lambda_{\ff_k}\}_{k=1}^{M-1}})
\;=\;\sum_{\gamma\in\mathbb{Z}_N}\;
|\{\lambda_k\}_{k=1}^{M-1},\gamma\rangle\,\langle\chi_\gamma |\;.
\end{equation}
Here $|\{\lambda_k\}_{k=1}^{M-1},\gamma\rangle$ are the
eigenvectors of $\widetilde{\mathbf{b}}$
(\ref{btilde},\ref{b-eigen}), and  $\chi_\gamma$ are some vectors,
such that
$\langle\chi_\gamma|\,\mathbf{Y}=\omega^\gamma\,\langle\chi_
\gamma|$, where $\mathbf{Y}$ is given by (\ref{Y}).

>From the other side, using the natural decomposition
(\ref{Q-decomp}) and (\ref{K-decomposition}), we should have for
the generic values of $f_k$
\begin{equation}\label{K-02}
\ds
\mathbf{K}^{(M-1)}(\{\lambda_{\ff_k}\}_{k=1}^{M-1})
\;=\;\sum_t\;|\Psi_t^{}\rangle
\; \left(\prod_{k=1}^{M-1}\,
q_{t}(\lambda_{\ff_n},\delta_{\ff_n})\right) \;
\langle\Psi_t^{(M-1)}|\;,
\end{equation}
where the summation is taken over all the eigenvalues $t$ of
$\tp$. Functions $q_t(\lambda,\delta)$ obey the Baxter equation
(\ref{simple-BE}) and due to proposition (\ref{prop-prop}) on page
(\pageref{prop-prop}) do not depend on $f_1,...,f_{M-1}$.
Comparing two decompositions (\ref{K-separation}) and (\ref{K-02})
at $f_k\mapsto 1$, we may conclude that the eigenvectors
$\langle\Psi_t^{(M-1)}|$ stick together in this limit
\begin{equation}
\ds\langle\Psi_t^{(M-1)}|\;\mapsto\;\langle\chi_\gamma|\;\;\;\forall\;t
=\{t_1,...,t_{M-1},t_M=(-\kappa)^M\omega^\gamma\}\;.
\end{equation}
So the functional Bethe ansatz formula
(\ref{q-sep}) follows from (\ref{K-separation}) and
(\ref{K-02}) up to a normalization. In addition, we know explicitly the
corresponding projector $\mathbf{K}^{(M-1)}$.

Detailed investigation shows that the explicit form of
$\langle\chi_\gamma|$ depends on a limiting procedure $f_n\mapsto
1$.  The reason is that when $f_n\mapsto 1$, $n=1,...,g$, then
$\tau^{(g)}_{m}\mapsto 0$, $m=0,...,g-1$. This follows from
the recursive definition of $h^{(g)}$ (\ref{h-definition}).
The proof is based on Proposition \ref{prop-zeros} on page
\pageref{prop-zeros} of the Appendix.
Because of the parameterization of
$u_m^{(n)}$ and $p_{j,m}^{(n)}$ contains the ratios of
$\tau^{(n)}_m$, the ambiguity arises. Some of $u_m^{(n)}$ and
$p_{j,m}^{(n)}$ of the form $0/0$ depends on the details of the
limiting procedure.

\subsection{Limiting procedure}

In this subsection we describe one of the possible limiting
procedures. Most simple $\mathbf{K}^{(M-1)}(\{\lambda_k\})$
appears when we choose $u^{(M-1)}_m=0$ for $m=1...M-1$. Let us
consider the set of infinitely small numbers,
\begin{equation}
\ds\varepsilon_k\,\mapsto\,0\;,\;\;\; k=1...M-1\;,
\end{equation}
such that any ratio $\varepsilon_n/\varepsilon_m\neq 1$ is
finite, and
\begin{equation}
\ds u^{(M-1)}_m\;=\;-\omega^{-1/2}\, {\varepsilon_{M-m}\over
\delta_1....\delta_{M-1}}\;,\;\;\;m=1...M-1\;,
\end{equation}
\begin{equation}
\ds u^{(M-1)}_M\;=\;-\omega^{-1/2}\,
{(\delta_1...\delta_{M-1})^{M-1} \over
\varepsilon_1...\varepsilon_{M-1}}\;.
\end{equation}
Recursion (\ref{ff-recursion},\ref{uw-prime}) with
\begin{equation}
\ds\kappa_{\phi_n}\;=\;{\kappa\over\delta_n}
\end{equation}
yields
\begin{equation}
\ds u_m^{(n)}\;=\;-\omega^{-1/2}\, {\varepsilon_{n-m+1}\over
\delta_1...\delta_n}\,+\,o(\varepsilon_1)\;,\;\;\; m=1...n
\end{equation}
while for $m=n+1...M-1$ all $u_m^{(n)}$ are regular. This
gives
\begin{equation}
\ds\tau^{(n)}_m\;=\;\tau^{(n)}_n\;
{\varepsilon_1...\varepsilon_{n-m} \over
(\delta_1...\delta_n)^{n-m}}\,+\,o(\varepsilon_1^{n-m})\;,\;
\;\;
m=0...n\;.
\end{equation}
In the limit $\varepsilon_k = 0$ we may choose
\begin{equation}
\ds {\tau^{(n)}_n\over \theta^{(n)}_n}\;=\;
\prod_{k=1}^n\;\omega^{-1/2}{\delta_k\over \lambda_k w_k}\;,
\end{equation}
where $w_k$ are given by (\ref{w-phi-m}), and
\begin{equation}\label{simple-theta}
\ds \theta^{(n)}_m\;=\;
(\omega^{1/2}\kappa)^{(n-m)(n-m-1)/2}
{\theta^{(n)}_n\over (\delta_1...\delta_n)^{n-m}}\;.
\end{equation}
$N$th powers of these formulas follow from Proposition
\ref{prop-prime} on page (\pageref{prop-prime}) of the Appendix.
Eq. (\ref{simple-theta}) gives
\begin{equation}\label{w/w}
\ds {w_{m+1}^{(n)}\over
w_{m}^{(n)}}\;=\;\omega^{1/2}\kappa\;,\;\;\; m=0...n-1\;.
\end{equation}

In the limit when $\epsilon_k\mapsto 0$ the regular $p_{j,m}^{(n)}$, $j=1,2,3$,
in $\Q_{\ff_n}$ (\ref{Q-elements}), entering the decomposition of
$\mathbf{K}^{(M-1)}$, are given by (\ref{xy-explicit}).
Below we will list the singular $p_{j,m}^{(n)}$, using notations
(\ref{singular-p},\ref{singular-w}).
\begin{itemize}
\item \underline{$m=1...n-1$}:
\begin{equation}
\ds p^{(n)}_{3,m}\;=\;q_1\;,\;\;\;
p^{(n)}_{2,m}\;=\;O(p^{(n)}_{1,m})\;,\;\;\;
x_{1,m}^{(n)}\;=\; {1\over
\omega^{1/2}\kappa}\,{\varepsilon_{n-m+1}\over\varepsilon_{n
-m}}\;.
\end{equation}
\item \underline{$m=n$}:
\begin{equation}
\ds p_{1,n}^{(n)}=q_0\;,\;\;\;
p_{2,n}^{(n)}=q_\infty\;\;\;\forall\;n
\end{equation}
and $p_{3,n}^{(n)}$ are regular, except
\begin{equation}
\ds p_{3,M-1}^{(M-1)}=q_\infty\;.
\end{equation}
\item \underline{$m=n+1...M-2$}:
All $p_{j,m}^{(n)}$ in this region are regular.
\item \underline{$m=M-1$}:
$p_{1,M-1}^{(n)}$ are regular except
\begin{equation}
\ds p_{1,M-1}^{(M-1)}=q_0\;,
\end{equation}
and
\begin{equation}
\ds
p_{2,M-1}^{(n)}=p_{3,M-1}^{(n)}=q_\infty\;\;\;\forall\;n\;.
\end{equation}
\item \underline{$m=M$}:
\begin{equation}
\ds p_{1,M}^{(n)}=q_\infty\;,\;\;\; p_{2,M}^{(n)} =
p_{3,M}^{(n)}=q_0\;\;\;\forall\;n.
\end{equation}
\end{itemize}
Substituting these expressions into (\ref{Q-elements}), one
obtains the following form of $N$th modified $Q$-matrix, $n\neq
M-1$
\begin{equation}\label{Q-n}
\ds\begin{array}{l} \ds
\langle\alpha|\mathbf{Q}^{(n)}|\beta\rangle\;=\;
{1\over\Phi(\beta_1+\beta_M-\alpha_M)}\,\prod_{m=1}^{n-1}\,
\delta_{\alpha_m,\beta_{m+1}}\times\\
\\
\ds
\Phi(\alpha_n)\,w_{Op^{(n)}_{3,n}}(\alpha_n-\beta_{n+1})\times\\
\\
\ds \prod_{m=n+1}^{M-2}\,\omega^{(\alpha_m-\beta_m)\beta_{m+1}}\,
{w_{p_{1,m}^{(n)}}(\beta_m-\alpha_m)\,
w_{p_{2,m}^{(n)}}(\beta_{m+1}-\beta_m)\over
w_{p_{3,m}^{(n)}}(\beta_{m+1}-\alpha_m)}\times\\
\\
\ds
{\Phi(\alpha_{M-1})\over\Phi(\beta_{M-1})}\,
w_{p_{1,M-1}^{(n)}}(\beta_{M-1}-\alpha_{M-1})\;,
\end{array}
\end{equation}
and for the last $(M-1)$-th $Q$-matrix
\begin{equation}\label{Q-last}
\ds\langle\alpha|\mathbf{Q}^{(M-1)}|\beta\rangle\;=\;
{1\over\Phi(\beta_1+\beta_M-\alpha_M)}\,\prod_{m=1}^{M-2}\,
\delta_{\alpha_m,\beta_{m+1}}\,
\Phi(\alpha_{M-1})\;.
\end{equation}
Here $\Phi$ is given by (\ref{Phi}). Explicit form of the
modified $Q$-operators (\ref{Q-n}) and (\ref{Q-last}) allows one
to prove the following
\begin{prop}
\begin{equation}
\ds
\mathbf{K}^{(M-1)}(\{\lambda_k\}_{k=1}^{M-1})\,(\zop_{m}-\zop_{m+1})\;=\;0\;,\;\;
\;m\,\in\,\mathbb{Z}_M\;.
\end{equation}
\end{prop}
\noindent\emph{Proof}. The matrix elements of $\Q^{(n)}$ depend on
$\beta_M-\alpha_M$, therefore
$\Q^{(n)}\,\zop_M\;=\;\zop_M\,\Q^{(n)}$ $\forall n$. The set of
delta-symbols in each $\Q^{(n)}$ gives
$\Q^{(n)}\,\zop_m\;=\;\zop_{m-1}\,\Q^{(n)}$, $m=2,...,n+1$. Also
the matrix elements of each $\Q^{(n)}$ depend on
$\beta_1+\beta_M$, hence $\Q^{(n)}\,(\zop_1-\zop_M)\;=\;0$
$\forall n$. Therefore $\forall m$
\begin{equation}
\ds \mathbf{K}^{(M-1)}\,(\zop_m-\zop_M)\,=\,
\Q^{(1)}...\Q^{(M-m)}\,(\zop_1-\zop_M)\,\Q^{(M-m+1)}...\Q^{(M-1)}\,=\,0.
\end{equation}
\hfill $\Box$

So, in our limiting procedure, $\chi_\gamma$ is defined by
\begin{equation}
\ds \langle\chi_\gamma|\,(\zop_m-\zop_{m+1})\;=\;0\;,\;\;\;
\langle\chi_\gamma|\,\mathbf{Y}\;=\;\omega^\gamma\,\langle\chi_\gamma|\;.
\end{equation}
Hence the matrix element of $\chi_\gamma$ is
\begin{equation}
\ds \langle\chi_\gamma|\alpha\rangle\;=\;
\chi_\gamma(\overline{\alpha})
\;\equiv\;\chi_\gamma(\overline{\alpha}\mod N),
\end{equation}
where
\begin{equation}
\ds\overline{\alpha}\;\stackrel{def}{=}
\;\sum_{m\in\,\mathbb{Z}_M}\,\alpha_m\;.
\end{equation}
Discrete function $\chi_\gamma(\overline{\alpha})$ is defined by
\begin{equation}\label{chi-eq}
\ds
{\chi_\gamma(\overline{\alpha}+M)\over
\chi_\gamma(\overline{\alpha})}\,=\,
(-\omega^{-1/2})^M\,\omega^{\gamma-\overline{\alpha}}\;.
\end{equation}
Solution of (\ref{chi-eq}) depends strongly on $M\mod N$.

\section{Discussion}

In this paper we investigated one of the simplest integrable
model, associated with the local Weyl algebra at the root of
unity. All such models always contain a classical discrete dynamic
of parameters. Nontrivial solution of this classical part of the model
provides the solution of the isospectrality problem of the spin
part. Well known nowadays results by Sklyanin, Kuznetsov
at al reveal that in the classical limit of the usual Toda chain (and
many other models) Baxter's quantum $Q$-operator is related to
the B\"acklund transformation of the classical system, see e.g.
\cite{Sklyanin-sep,Sklyanin-rev,KS-manybody,Sklyanin-fun,kss-dst}
etc. In our case we have the B\"acklund transformation of the
classical counterpart and modified $Q$-operator in the quantum
space simultaneously. It is unusual that solving the quantum
isospectrality problem, we miss the commutativity of the modified
$Q$-operators.

Our results are the
explicit construction of $(M-1)$-parametric family of quantum
inhomogeneous transfer matrices with the same spectrum as the
initial homogeneous one and the explicit construction of the
corresponding similarity operator (\ref{K-operator}). We hope that the
solution of the isospectrality problem will help to solve the
model with arbitrary $N$ in the thermodynamic limit exactly.

As one particular application of the isospectrality we have
obtained the quantum separation of variables. Previously, there was
a hypothesis formulated for the usual quantum Toda chain, that
the product of \emph{operators} $Q$, taken in the special points,
is related to the functional Bethe ansatz. In this paper we have
established it explicitly, but for the product of \emph{modified
operators} $Q$.

Note in conclusion that this method may be applied to any model, based
on the local Weyl algebra. One may mention the chiral Potts model
\cite{CPM,BS-cpm} and the Zamolodchikov-Bazhanov-Baxter model in
the vertex formulation \cite{mss-vertex}. Moreover, all two
dimensional integrable models with the local Weyl algebra are
particular cases of the general three dimensional model, and their
classical counterparts are known \cite{Sergeev}.

{\bf Acknowledgements} The authors are grateful to R. Baxter, V.
Bazhanov, V. Mangazeev, G. Pronko, E. Sklyanin, A. Belavin, Yu.
Stroganov, A. Isaev, G. von Gehlen, V. Tarasov, P. Kulish, F.
Smirnov and R. Kashaev for useful discussions and comments.

\appendix

\section{Fay's identity in the rational limit}

It is well known what is the rational limit of $\Theta$-function
on a generic algebraic curve. $\Theta$-function by definition is
\begin{equation}\label{Theta-function}
\ds\Theta^{(g)}(\mathbf{z})\;=\;\sum_{\mathbf{n}\in\mathbb{Z}^g}\;
\EXP^{i\pi(\mathbf{n},\Omega\mathbf{n})\,+\,2 i \pi
(\mathbf{n},\mathbf{z})}\;,
\end{equation}
where $\mathbf{z}=(z_1,...,z_g)\in\mathbb{C}^g$ and the imaginary
part of the symmetrical period matrix $\Omega$ is positively
defined. $\Theta$-function may be defined recursively as
\begin{equation}\label{Theta-function-recursion}
\ds \Theta^{(g)}(\mathbf{z})\;=\; \sum_{n_g\in\mathbb{Z}}\;
\Theta^{(g-1)}(\mathbf{z}'+\Omega_g'n_g^{})\,
\EXP^{i\pi\Omega_{g,g}^{}n_g^2\,+\, 2 i \pi n_h^{} z_g^{}}\;,
\end{equation}
where $\mathbf{z}'=(z_1,...,z_{g-1})$ and
$\Omega_g'=(\Omega_{1,g}^{},...,\Omega_{g-1,g}^{})$.

The rational limit corresponds to
\begin{equation}
\ds \EXP^{i\pi\Omega_{n,n}+2 i \pi z_n}\;=\;-f_n^{}\;,
\end{equation}
and
\begin{equation}
\ds \EXP^{i\pi\Omega_{n,n}}\;\mapsto\;0\;,\;\;\;
\EXP^{i\pi\Omega_{k,n}}\;\mapsto\; {(q_k-q_n)\,(p_k-p_n)\over
(q_k-p_n)\,(p_k-q_n)}\;.
\end{equation}
Explicit details of the limiting procedure are rather tedious. We
should better invent all the necessary components in the rational
limit directly and formulate the rational limit of the Fay
identity. The proof of all the propositions is analogous to that
in the algebraic geometry with the elementary replacement of the
usual $\Theta$-function methods to the analysis of the polynomial
structure and intensive use of the Main Algebraic Theorem.

\subsection{Rational limit of arbitrary algebraic curve jacobian}

Let $\{p_n,q_n\}_{n=1}^g$ be a set of $2g$ complex parameters.
``$\Theta$-function'' in the rational
limit is a function of $g$ free complex arguments $f_1,...,f_g$,
the set $\{p_n,q_n\}_{n=1}^g$ is its parameters.
``$\Theta$-function'' may be defined recursively with respect to
$g$ as follows
\begin{equation}
\ds\begin{array}{l} \ds \Theta^{(0)}\;=\;1\;,\;\;\;
\Theta^{(1)}(f_1)\;=\;1\,-\,f_1\;,\\
\\
\ds \Theta^{(2)}(f_1,f_2)\;=\; 1-f_1- f_2 + d_{1,2} f_1 f_2\;,
\end{array}\end{equation}
and so on,
\begin{equation}\label{Theta-fun}
\ds \Theta^{(g)}(\{f_n\}_{n=1}^g)\,=\,
\Theta^{(g-1)}(\{f_n\}_{n=1}^{g-1})\,-\, f_g\,\Theta^{(g-1)}(\{f_n
d_{n,g}\}_{n=1}^{g-1})\;.
\end{equation}
The recursion follows from (\ref{Theta-function-recursion}), where
$n_g=0,1$. Parameters $\{p_n,q_n\}_{n=1}^g$ enter the definition
of the phase shifts
\begin{equation}\label{alg-phase-shift}
\ds d_{n,m}\;=\; {(p_n-p_m)\,(q_n-q_m)\over
(p_n-q_m)\,(q_n-p_m)}\;.
\end{equation}
It is useful to introduce function $\sigma_n(z)$
\begin{equation}\label{S-fun}
\ds \sigma_n(z)\;=\; {p_n\,-\,z\over q_n\,-\,z}\;.
\end{equation}
Let  by definition
\begin{equation}
\ds H^{(g)}(\{f_n\}_{n=1}^g)\;=\; \Theta^{(g)}(\{f_n\prod_{m\neq
n}\sigma_n(q_m)\}_{n=1}^g)\;.
\end{equation}
Recursion definition of $H^{(g)}$ is
\begin{equation}\label{H-recursion}
\ds H^{(g)}(\{f_n\}_{n=1}^g)\;=\;
H^{(g-1)}(\{f_n\sigma_n(q_g)\}_{n=1}^{g-1}) \,-\,
f_g\prod_{m=1}^{g-1}\sigma_g(q_m)\,
H^{(g-1)}(\{f_n\sigma_n(p_g)\}_{n=1}^{g-1})\;,
\end{equation}
where $\sigma_n(q_g) d_{n,g}=\sigma_n(p_g)$ is taken into account.

\begin{prop}\label{prop-zeros} (Zeros of $\Theta$-function)
\begin{equation}
\ds H^{(g)}(\{f_n\}_{n=1}^g)\,=\,0\;\;\Longleftrightarrow\;\;
f_n\,=\,\prod_{m=1}^{g-1}\sigma_n(z_m)^{-1}\;,
\end{equation}
where $z_m\in\mathbb{C}$, $m=1,...,g-1$.
\end{prop}
\noindent\emph{Proof}. We will use the mathematical induction method. This
proposition is obviously valid for $g=1$. Suppose, one has
$H^{(g-1)}(\{\prod_{m=1}^{g-2}\sigma_n(z_m)^{-1}\}_{n=1}^{g-1})=0$
$\forall$ $z_1...z_m$, $p_1...p_{g-1}$, $q_1...q_{g-1}$ for some
$g$. Consider now $H^{(g)}$. Due to recursion relation
(\ref{H-recursion})
\begin{equation}
\ds H^{(g)}(\{\prod_{m=1}^{g-1}\sigma_n(z_m)^{-1}\}_{n=1}^g)\;=\;
{P_g(z_1,...,z_m)\over
\prod_{n<n'}(q_n-q_{n'})\,\prod_{n=1}^g\prod_{m=1}^{g-1}(p_n-z_m)}\;,
\end{equation}
where $P_g$ is a polynomial of the power $g$ with respect to each
$z_m$.

Consider the case when $z_{g-1}=q_g$, so
$\prod_{m}\sigma_g(z_m)^{-1}=0$. From recursion relation
(\ref{H-recursion}) and the supposition of the induction it
follows that $H^{(g)}=0$, so due to the symmetry of $H^{(g)}$ we
obtain
\begin{equation}
\ds H^{(g)}(\{\prod_{m=1}^{g-1}\sigma_n(z_m)^{-1}\}_{n=1}^{g-1})
\,=\,0\;\;\;\textrm{if}\;\;\; z_m=q_n\;\;\forall\;n,m\;.
\end{equation}
Next consider the case when $z_{g-1}\mapsto p_g$, hence
$\prod\sigma_n(z_m)\sim (p_g-z_{g-1})^{-1}\mapsto\infty$. Due to
recursion relation (\ref{H-recursion}) and the supposition of the
induction it follows that $H^{(g)}$ is regular in $z_{g-1}=p_g$.
Again, due to the symmetry of $H^{(g)}$
\begin{equation}
\ds \textrm{Res}_{z_m=p_n}
H^{(g)}(\{\prod_{m=1}^{g-1}\sigma_n(z_m)^{-1}\}_{n=1}^g) \;=\; 0
\;\;\;\forall\;\;n,m\;.
\end{equation}
Therefore,
\begin{equation}
\ds P_g(z_1...z_{g-1})\;\sim\; \prod_{n=1}^g\prod_{m=1}^{g-1}\,
(p_n-z_m)\, (q_n-z_m)\;.
\end{equation}
This product is the polynomial of the power $2g$ with respect to
each $z_m$, and this contradicts with the structure of $P_g$, therefore
$P_g\equiv 0$. Hence we prove
\begin{equation}
\ds
H^{(g)}(\{\prod_{m=1}^{g-1}\sigma_n(z_m)^{-1}\}_{n=1}^g)\;=\;0\;\;\;
\forall\;\;z_1,...,z_{g-1}\;.
\end{equation}

Backward, consider equation $H^{(g)}(\{f_n\}_{n=1}^g)=0$. Since
$H^{(g)}$ is the first power polynomial with respect to each
$f_n$, solution of $H^{(g)}=0$ is a rational
$\mathbb{C}^{g-1}$ variety, so as
$f_n=\prod_{m=1}^{g-1}\sigma_n(z_m)^{-1}$. \hfill $\Box$

\begin{prop}\label{prop-casor} (Casoratti determinant representation)
\begin{equation}\label{casor}
\ds H^{(g)}(\{f_n\}_{n=1}^g)\;=\;{\ds \det
|q_j^{i-1}\,-\,f_j^{}\,p_j^{i-1}|_{i,j=1}^g\over
\ds\prod_{i>j}\,(q_i^{}\,-\,q_j^{})}\;.
\end{equation}
\end{prop}
\noindent\emph{Proof}. Evidently, zeros of the right hand side of (\ref{casor})
correspond to the case when the columns of the matrix
$|q_j^{i-1}\,-\,f_j^{}\,p_j^{i-1}|_{i,j=1}^g$ are linearly
dependent, i.e. $\exists$ $c_i$, $i=1,...,g$:
\begin{equation}
\ds
\sum_{i=1}^g\,c_i^{}\,q_j^{i-1}
\;=\; f_j^{}\sum_{i=1}^g\,c_i^{}\,p_j^{i-1}\;.
\end{equation}
Let $\ds \sum_{i=1}^g\,c_i^{}\,z^{i-1}\;=\;\prod_{m=1}^{g-1}\,(z-z_m)$, then
the determinant is zero if and only if
\begin{equation}
\ds f_j^{}\;=\;\prod_{m=1}^{g-1}\,{q_j^{}-z_m^{}\over p_j^{}-z_m^{}}
\;=\; \prod_{m=1}^{g-1}\sigma_j^{}(z_m^{})^{-1}\;.
\end{equation}
Therefore, due to proposition (\ref{prop-zeros}) the determinant
is proportional to $H^{(g)}$. The denominator in the right hand
side of (\ref{casor}) is the subject of normalization
when $f_j=0$.\hfill  $\Box$

\begin{prop}\label{prop-prime} (Theta-functions and prime forms)
\begin{equation}
\ds H^{(g)}(\{f_n = \sigma_n(\lambda) \prod_{m=1}^g
\sigma_n(z_m)^{-1}\}_{n=1}^g)\;=\;\left(\prod_{m=1}^g\,
{\lambda-z_m\over \lambda-q_m}\right)\; H^{(g)}(\{f_n =
\prod_{m=1}^g \sigma_n(z_m)^{-1}\}_{n=1}^g)\;.
\end{equation}
\end{prop}
\noindent\emph{Proof}. A simple application of the recursion
relation (\ref{H-recursion}) to $H^{(g+1)}(\{...\})=0$, where
$p_{g+1}=\infty$ and $q_{g+1}=\lambda$.\hfill $\Box$

\subsection{The Fay identity}

\begin{prop}\label{prop-fay} (Fay's identity)
\begin{equation}\label{fay-identity}
\ds\begin{array}{l} \ds (A-D)(C-B)\,
H^{(g)}(\{f_n{\sigma_n(A)\over\sigma_n(B)}\}_{n=1}^g)\,
H^{(g)}(\{f_n{\sigma_n(C)\over\sigma_n(D)}\}_{n=1}^g)\;+ \\
\\
\ds (A-B)(D-C)\,
H^{(g)}(\{f_n{\sigma_n(A)\over\sigma_n(D)}\}_{n=1}^g)\,
H^{(g)}(\{f_n{\sigma_n(C)\over\sigma_n(B)}\}_{n=1}^g)\;= \\
\\
\ds (A-C)(D-B)\, H^{(g)}(\{f_n\}_{n=1}^g)\,
H^{(g)}(\{f_n{\sigma_n(A)\sigma_n(C)
\over\sigma_n(B)\sigma_n(D)}\}_{n=1}^g) \;.
\end{array}
\end{equation}
\end{prop}
\noindent\emph{Proof}. Let $LHS(\{f_n\}_{n=1}^g)$ be the left hand side of
(\ref{fay-identity}). It is a second order polynomial with respect
to each of $f_1,...,f_g$. Using proposition (\ref{prop-prime}), it
may be shown that
\begin{equation}
\begin{array}{l}
\ds LHS(\{f_n={\sigma_n(B)\sigma_n(D)\over\sigma_n(A)\sigma_n(C)}
\prod_{m=1}^{g-1}\sigma_n(z_m)^{-1}\}_{n=1}^g)\;=\\
\\
\ds LHS(\{f_n=
\prod_{m=1}^{g-1}\sigma_n(z_m)^{-1}\}_{n=1}^g)\;=\;0
\end{array}
\end{equation}
for any set of $z_1,...,z_{g-1}$. Due to Proposition
(\ref{prop-zeros}), it means unambiguously
\begin{equation}
\ds LHS(\{f_n\})\;=\;\textrm{const}\; H^{(g)}(\{f_n\}_{n=1}^g) \;
H^{(g)}(\{f_n{\sigma_n(A)\sigma_n(C)\over
\sigma_n(B)\sigma_n(D)}\}_{n=1}^g)\;.
\end{equation}
Taking, for example, $\ds f_n={\sigma_n(D)\over\sigma_n(A)}
\prod_{m=1}^{g-1}\sigma_n(z_m)^{-1}$, we fix the constant and
obtain finally the given form (\ref{fay-identity}) of the
coefficients.\hfill $\Box$

\subsection{Application to the relativistic Toda chain}

Let all parameters $\{p_n,q_n\}_{n=1}^\infty$ be related by
\begin{equation}\label{hyper-def}
\ds  p_n\,+\,q_n\;=\;1\,-\,\kappa^N p_n q_n\;\;\;\forall\;n\;,
\end{equation}
so that
\begin{equation}\label{hyper-phi}
\ds p_n\;=\;{1\over \Om_{\ff_n}^{}}\;,\;\;\;
q_n\;=\;{1\over\Om_{\ff_n}^*}
\end{equation}
in the notations of (\ref{phi-parametrization}). Note,
\begin{equation}
\ds \sigma_n(-1/\kappa^N)\;=\;\sigma_n(0)\,\sigma_n(1)\;,
\end{equation}
and in the $\Om,\Om^*$-notations (\ref{s-function})
\begin{equation}
\ds \EXP^{2 i \phi_n}\;=\;\sigma_n(0)^{-1}\;,\;\;\;
s_{n}(1)\;=\;\sigma_n(1)\;,\;\;\;
s_{n}(\Om_m^*)\;=\;\sigma_n(q_m)\;,\;\;\;
s_n(\Om_m)\;=\;\sigma_n(p_m)\;.
\end{equation}
Besides, when $\{p_n,q_n\}$ obey (\ref{hyper-phi}), one has
$H^{(g)}\equiv h^{(g)}$ (see (\ref{h-definition})).

Taking $z_1=...=z_m=1$ and $z_{m+1}=...=z_{g-1}=\infty$ in
proposition \ref{prop-zeros} on page \pageref{prop-zeros}, one
obtains
\begin{equation}\label{zeros-h}
\ds h^{(g)}(\{\EXP^{2i
m\phi_n}\}_{n=1}^g)\;=\;0\;,\;\;\;m=0,...,g-1\;.
\end{equation}

Consider discrete equations  (\ref{triplet}) and (\ref{theta-rec})
with $m\in \mathbb{Z}$ and $n\in\mathbb{Z}_+$.
Omitting the index $n$ in
$\tau$-functions, we rewrite these relations as follows
\begin{equation}\label{triplet-fay}
\begin{array}{l}
\ds \left(\delta_g^{}\delta_g^{*}\, \tau^{\prime}_{m}\,
\tau^{}_m\, \theta^{}_{m-1}\right)^N \,=\, \left(\delta_{g}^{}\,
\tau^{\prime}_{m-1}\, \tau^{}_{m}\,
\theta^{}_{m}\right)^N \\
\\
\ds +\, \left(\delta_{g}^{*}\, \tau^{\prime}_{m+1}\,
\tau^{}_{m-1}\, \theta^{}_{m-1}\right)^N \,+\, \left(\kappa\,
\tau^{\prime}_m\, \tau^{}_{m-1}\, \theta^{}_m\right)^N\;
\end{array}
\end{equation}
and
\begin{equation}\label{theta-rec-fay}
\ds \left(\theta^{\prime}_m\right)^N\;=\;
\left({w_{\ff}\over\tau^{}_{m-1}}\right)^N\,
\left(\left(\theta^{}_m\tau^{\prime}_{m-1}\right)^N\,-\,
\left(\delta_{g}^{*}\theta^{}_{m-1}
\tau^{\prime}_{m}\right)^N\right)\;.
\end{equation}

\begin{prop}\label{prop-baecklund}
Let
\begin{equation}\label{tt-H}
\ds (\tau_m)^N\;=\;H^{(g-1)}(\{f_n\EXP^{2i\phi_n m}
\}_{n=1}^{g-1})\;,\;\;\; (\theta_m)^N\;=\; H^{(g-1)}(\{f_n\EXP^{2
i\phi_n m}\sigma_n(1)\}_{n=1}^{g-1})\;.
\end{equation}
Then the complete solution of equations (\ref{triplet-fay}) and
(\ref{theta-rec-fay}) are
\begin{equation}\label{tt-H-2}
\ds (\tau_m')^N\;=\;H^{(g)}(\{f_n\EXP^{2i\phi_n
m}\}_{n=1}^{g})\;,\;\;\; (\theta_m')^N\;=\; H^{(g-1)}(\{f_n\EXP^{2
i\phi_n m}\sigma_n(1)\}_{n=1}^{g})\;,
\end{equation}
where $\EXP^{2i\phi_g}=(\delta_g/\delta^*_g)^N$, all
$f_1,...,f_{g-1}$ are the same in (\ref{tt-H}) and (\ref{tt-H-2}),
and $f_g$ is arbitrary extra complex number.
\end{prop}
Proof. Index $m$ enters (\ref{tt-H}) as a scale of $f_n$, so
later we may consider only $m=0$.

Note that due to (\ref{H-recursion}) and the fact
$\sigma_g(q_g)^{-1}=0$, one has
\begin{equation}
\ds H^{(g)}(\{f_n\sigma_n(q_g)^{-1}\}_{n=1}^g)\;\equiv\;
H^{(g-1)}(\{f_n\}_{n=1}^{g-1})\;.
\end{equation}

For given set of $\{f_n\}_{n=1}^g$ consider two copies of Fay's
identity: the first one with
\begin{equation}
\ds A=0\,,\;\; B=q_g\,,\;\; C=1\,,\;\; D=-{1\over\kappa^N}\;,
\end{equation}
and the second one with
\begin{equation}
\ds A=\infty\,,\;\; B=q_g\,,\;\; C=-{1\over\kappa^N}\,,\;\; D=1\;.
\end{equation}
Condition (\ref{hyper-def}) is implied. Excluding $\ds
H^{(g)}(\{f_n\sigma_n(1)^{-1}\}_{n=1}^g)$ from both these
equation, one obtains exactly (\ref{triplet-fay}) with the
identification (\ref{tt-H},\ref{tt-H-2}). Equation
(\ref{theta-rec-fay}) corresponds to the Fay identity with
\begin{equation}
\ds A= 0\,,\;\; B = q_g \,,\;\; C= 1\,,\;\; D = \infty\,.
\end{equation}
\phantom{.} \hfill $\Box$

Note in conclusion, evolution equation (\ref{Hirota})
corresponds to the Fay identity with
\begin{equation}
\ds A = \infty\,,\;\; B = 0 \,,\;\; C= -{1\over\kappa^N}\,,\;\;
D = 1\,.
\end{equation}

%\section{$N=2$ case}

%\tableofcontents

\bibliographystyle{amsplain}

\end{document}